\documentclass[12pt]{article}
\usepackage{amsthm}
\usepackage{amsmath}
\usepackage{natbib}
\usepackage[colorlinks,citecolor=blue,urlcolor=blue,filecolor=blue,backref=page]{hyperref}
\usepackage{graphicx}
\usepackage[english]{babel}
\usepackage[utf8]{inputenc}
\usepackage{amsmath}
\usepackage{amssymb}
\usepackage{graphicx}
\usepackage{hyperref}
\usepackage{graphicx}
\usepackage{comment}
\usepackage{mathrsfs}
\usepackage{amsthm}
\usepackage{bbm}
\usepackage{rotating}
\usepackage{lscape}
\usepackage[colorinlistoftodos]{todonotes}
\usepackage{lipsum}
\usepackage{comment}
\usepackage{natbib}
\usepackage[colorlinks,citecolor=blue,urlcolor=blue,filecolor=blue,backref=page]{hyperref}
\usepackage[algoruled,boxed,lined]{algorithm2e}

\usepackage{graphicx}
\newtheorem{theorem}{Theorem}

\newcommand{\blind}{0}

\begin{document}

\def\spacingset#1{\renewcommand{\baselinestretch}%
{#1}\small\normalsize} \spacingset{1}


\if0\blind
{
  \title{\bf {Spatial shrinkage via the product independent Gaussian process prior}}
  \author{Arkaprava Roy$^{1}$, 
       Brian J. Reich$^{2}$,
       Joseph Guinness$^{3}$,\\
       {Russell T. Shinohara$^{4}$, Ana-Maria Staicu$^{2}$}\\
       $^{1}$University of Florida, $^{2}$North Carolina State University,$^{3}$Cornell University,\\
       $^{4}$University of Pennsylvania}
  \maketitle
} \fi

\if1\blind
{
  \bigskip
  \bigskip
  \bigskip
  \begin{center}
    {\LARGE\bf }
\end{center}
  \medskip
} \fi

\bigskip
\begin{abstract}
We study the problem of sparse signal detection on a spatial domain. We propose a novel approach to model continuous signals that are sparse and piecewise-smooth as the product of independent Gaussian processes (PING) with a smooth covariance kernel. The smoothness of the PING process is ensured by the smoothness of the covariance kernels of the Gaussian components in the product, and sparsity is controlled by the number of components. The bivariate kurtosis of the PING process implies that more components in the product results in the thicker tail and sharper peak at zero. We develop efficient computation algorithm based on spectral methods. The simulation results demonstrate superior estimation using the PING prior over Gaussian process (GP) prior for different image regressions. We apply our method to a longitudinal MRI dataset to detect the regions that are affected by multiple sclerosis (MS). computation in this domain.
\end{abstract}

\noindent%

\spacingset{1.45}

\maketitle

\section{Introduction}
In this paper, we discuss linear regression models for two or three-dimensional image responses, image covariates, or both, in which the signal is assumed to be continuous, sparse and piecewise smooth. The methodological development is motivated by a study of multiple sclerosis using magnetic resonance imaging \citep{sweeney,pomann,mejia}, where subjects with multiple sclerosis (MS) are imaged repeatedly over multiple hospital visits, and the objective is to identify the brain regions that are damaged over time. Although a healthy brain would not change much during the study period, a diseased brain is expected to exhibit changes in a small number of regions of interest that are associated with the disease. This is an example of image-on-scalar regression, in which the signal is desired to be continuous, sparse and piecewise smooth. 

Modeling a continuous, sparse and piecewise-smooth signal for high-dimensional data poses several challenges such as 
1) complex spatial dependence of the data, 2) accounting for a simultaneously sparse and continuous; here sparsity is defined in terms of the number of non-zero smooth pieces that comprise the signal and 3) accommodating a very large dimensional signal. 

For sparse estimation, there are some traditional approaches that we discuss here. In a frequentist framework, there is lasso-type penalization \citep{lasso} but this cannot ensure smooth changes from zeros to non-zero subregions. Using fused lasso \citep{fused}, one can ensure both sparsity and smoothness in the estimation. None of these approaches allow for quantification of estimation uncertainty. In a Bayesian framework, parameter sparsity is modeled using the traditional spike and slab prior \citep{spikeslab}, the horseshoe prior \citep{Carvalho}, normal-gamma prior \citep{Griffin}, double-Pareto prior \citep{Armagan} or Dirichlet-Laplace prior \citep{Bhattacharya}. However, none of these priors ensures a smooth spatial structure. In the context of high dimensional data, this adds computational challenges as well. Reducing computational demand is one issue that we try to address in this paper.

We review some of the research on sparse and spatially smooth parameter estimation for different types of image regression. For image-on-scalar regression, \cite{SIC} and \cite{chen} tackled a similar problem. The first paper considers a Laplacian type penalty and the second paper introduces a fused SCAD type penalty to account simultaneously for spatial smoothness and sparsity. In the context of functional magnetic resonance imaging (fMRI) studies \cite{fmri1} and \cite{fmri2} too considered similar regression models. Their estimation approach uses a spike and slab prior to induce sparsity and considers spatial smoothness for the selection. However, the approach does not ensure that the estimated signal is smooth. In scalar-on-image regression, there is limited work on sparse and piece-wise smooth signal estimation. \cite{Goldsmith} and \cite{Li} proposed priors that account separately for spatial dependence and sparsity. For the same problem, \cite{Wang} proposed a penalty based on the total variation. Spatial dependence is still not fully incorporated in this approach. In \cite{kang}, the proposed soft-thresholded Gaussian process prior account for both spatial dependence and sparsity simultaneously. The method is computationally very expensive. In the context of image response and image predictors, there is a more limited research \citep{image1, image4, image5}. \cite{vcii1}, \cite{vcii2} considered varying co-efficient model that accounts for sparsity but not smoothness. 
To the best of our knowledge, only \cite{Boehm-Vock} and \cite{jhuang} consider both smoothness and sparsity for image-on-image regression. Their methodology captures the spatial dependence using copulas, which is computationally expensive for large datasets. Our prior has nice conjugacy structure which leads to computational advantages.  

In this paper, we propose a novel prior which can be used to estimate continuous, sparse and piecewise smooth functions. We construct the prior as the location-wise product of independent Gaussian processes with smooth covariance kernel. The proposed prior has both high mass around zero, which creates sparsity in the estimation, and a smooth covariance kernel that ensures large support for the spatially varying function. To handle the heavy computational burden associated with this kind of prior, we propose to use the discrete Fourier transformation (DFT) that decorrelates the stationary part of the response. Specifically, we use the fast Fourier transformation (FFT). The FFT algorithm requires regularly spaced input data. In reality, the datasets are not often on a regular grid. To bypass this issue, we propose a fast imputation technique to transform the data into a regular grid. If the dimension of the dataset is manageable for computation in the spatial domain, one can exploit the conjugacy structure of our prior to get the full conditional distribution of parameters given the error process is Gaussian. We analyze the performance of our prior with respect to commonly used Gaussian process (GP) prior in different linear image regressions with signals that are sparse, piecewise smooth and continuous.

We organize the remainder of the paper as follows. In the next section, we describe the image-on-scalar regression model along with the new sparse prior process. We discuss the usage of our new prior to other image regression models in Section~\ref{extend}. In Section~\ref{compute}, we describe other computational aspects that we use for faster computation. In Section~\ref{simulation}, we provide several simulation results to evaluate the performance of this new prior for different image regression models extensively. We apply our method to a longitudinal magnetic resonance imaging (MRI) data in Section~\ref{real} and end with some concluding remarks in Section~\ref{conclude}.

\section{The modeling framework}
\label{model}
Our research is motivated by a longitudinal study of multiple sclerosis (MS) via magnetic resonance imaging (MRI) images. We introduce the main ideas for the case when we have images collected at multiple time points for a single subject. Specifically, let $Y_i(v)$ be the intensity of $i$-$th$ MRI image collected at time $t_i$ and for a 3-dimensional voxel $v$ for an MS subject. Consider the following linear image-on-scalar regression model
\begin{align}
&Y_i(v) = \alpha(v) + t_i\beta(v) + E_i(v),\label{modelIS}
\end{align}
where $\alpha(v)$ is a spatially-varying intercept function and $\beta(v)$ is an unknown continuous function that quantifies the effect of time, it is assumed that $\beta(\cdot)$ is in addition piecewise smooth and sparse. The error $E_i(v)$ is a spatially-correlated mean-zero error process, independent across visits. 
We propose to model $\beta(v)$ as a product of independent Gaussian processes (GPs). We formally describe its properties and the error process in the remainder of this section. 
\subsection{PING process}
\label{PING}
Let $\beta_1(v),...,\beta_q(v)$ be $q$ independent and identically distributed GPs with mean $E\{\beta_j(v)\}=0$, variance $V\{\beta_j(v)\}=1$, and covariance kernel Cov\{$\beta_j(v),\beta_j(v')$\}=$K(v,v')$ for $j=1,\ldots,q$. The zero-mean \textbf{P}roduct of \textbf{IN}dependent \textbf{G}aussian (PING) stochastic process is defined as the point-wise product of independent Gaussian processes (GPs), $\beta(v) = \sigma\beta_{1}(v)\cdot \beta_{2}(v)\cdot\ldots\cdot \beta_{q}(v)$ where $\sigma>0$ is a scale parameter. The stochastic process $\beta = \{\beta(v) : v\in{\cal V}\}$ constructed in this way is denoted $\beta\sim\mbox{PING}(q,\sigma^2, K)$.

\subsubsection{Properties of the marginal distribution}

We first discuss the distribution of the PING process at a single location $v$. The theoretical properties of the marginal distribution of $\beta(v)$ have been studied by \cite{prodN} and \cite{gaunt2018products}. \cite{gaunt2018products} provides detailed results on characteristic function and propose estimates for the tail behavior of product normals. We briefly revise some of its properties here for completeness. The marginal density function $f_q(x)$ for the product of $q$ standard normals is given by $f_q(x)=\frac{1}{(2\pi)^{q/2}}G_{0,q}^{q,0}(\frac{x^2}{2^q}|0),$ where $G(\cdot)$ denotes the Meijer G-function \citep{prodN}. The $k^{th}$ marginal moment is $E\{\beta(v)^k\}=\big[(k-1)!!\big]^q$ where $n!!$ is the product of all numbers from 1 to $n$ that have the same parity as $n$. The density is unimodal and symmetric about zero; thus, all the odd-order moments are zero. The variance is $V\{\beta(v)\}=\sigma^2$. The marginal kurtosis is equal to $3^q-3$ which is an increasing function of $q$. As a result, the marginal density has thicker tail and sharper peak at zero for larger $q$. This is depicted in the first row of Figure~\ref{contour}. Furthermore, $f_q(x)\sim q\exp(-q(x^2/2^q)^{1/q})$ as $x$ goes to infinity and thus the tail is heavier than Gaussian for $q>1$ and it gets heavier as $q$ increases; (see \cite{gaunt2018products}). 

\subsubsection{Properties of the bivariate distribution}
Next, we study the bivariate properties of the PING process at a pair of locations $v_1$ and $v_2$. From the construction of the PING process with $q$ components, this bivariate distribution is in fact the distribution of the product of $q$ bivariate normals. Simple calculations show that its mean is $E\{\beta(v_j)\}=0$ for $j = 1, 2$, and its covariance is $\text{Cov}\{\beta(v_1),\beta(v_2)\} = \sigma^2K^q(v_1,v_2)$, implying a correlation coefficient that is smaller than the correlation of each individual Gaussian components and that further decays with the number of components, $q$. In particular, if $K(\cdot)$ is the powered exponential correlation kernel, $K(v_1, v_2)=\exp\{-\big(\frac{\|v_1-v_2\|_2}{\rho}\big)^\nu\}$, the PING covariance is $\exp\{-\big(\frac{\|v_1-v_2\|_2}{\rho q^{-1/\nu}}\big)^\nu\}$. Therefore, while the covariance decreases with $q$ for a fixed kernel function, strong spatial correlation can be maintained for large $q$ by simply increasing the parameter $\rho$ with $q$. The smoothness of the product process is the same as that of its individual components for these power exponential cases. We expect that separating sparsity and spatial dependence hold for other kernel functions as well. To quantify the shrinkage properties, we study the kurtosis of this product distribution. 
Kurtosis of a multivariate random variable $Z$ of dimension $p$ with mean $\mu_z$ and covariance matrix $\Sigma_Z$ is defined as $E[(Z-\mu_Z)^{T}\Sigma_{Z}^{-1}(Z-\mu_Z)]^2-p(p+2)$ \citep{Mardia}. The kurtosis of a general product of bivariate normal random variable is summarized by the following theorem.

\begin{figure}[htbp]
    \centering
    \includegraphics[width = 0.6\textwidth]{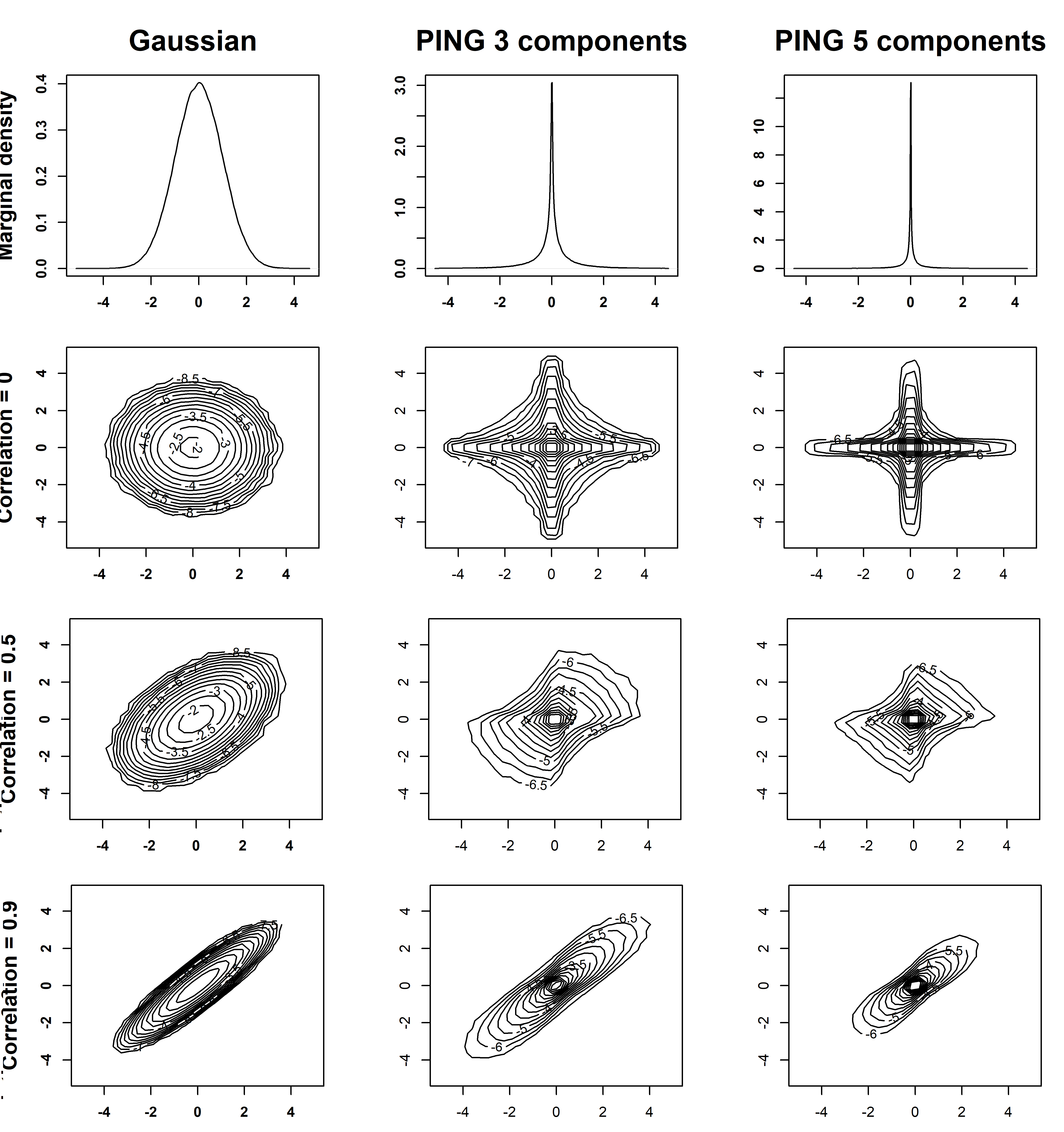}
    \caption{Comparison of Gaussian density with PING-3 and PING-5 densities.}
    \label{contour}
\end{figure}

\begin{theorem}
Let $Z_1,\ldots,Z_q$ be such that $Z_i \overset{ind}{\sim}$ $\mathrm{BVN}(0,0,\sigma_{i1}^2, \sigma_{i2}^2, \rho)$ for $i=1,\ldots, q$ and $P_q =Z_{1}\bigodot\ldots\bigodot Z_{q}$. The mean and the covariance matrix of $P_q$ are $E(P_q) = 0$ and
\[
\mathrm{Cov}(P_q)=
\left[ {\begin{array}{cc}
    \prod_{i=1}^q \sigma_{i1}^2 & \rho^q\prod_{i=1}^q \sigma_{i1}\sigma_{i2} \\
    \rho^q\prod_{i=1}^q \sigma_{i1}\sigma_{i2} &  \prod_{i=1}^q \sigma_{i2}^2 \\
    \end{array} } \right].
\] The kurtosis is ${\mathrm{Kurt}}(q,m) = \frac{2\times 3^q}{(1-m^{q})^2}[1+2(\frac{(1+2m)m}{3})^q + (\frac{1+2m}{3})^q - 4m^{q}]-8$, with $m=\rho^2$ and it increases with $q$.
\label{bvt}
\end{theorem}
Here ``BVN'' stands for bivariate normal and $\bigodot$ denotes the element wise product. We allow varying variances for individual components as the kurtosis does not depend on the variances. See Supplementary materials for details.
The ``increasing'' property of the kurtosis results from the application of arithmetic and geometric means inequality. The distribution of $\beta(v)$ for two locations $v_1$ and $v_2$ under PING process has the above mentioned properties with $\sigma_{i1}=\sigma_{i2}=1$ for $i>1$. Since it is a unimodal symmetric distribution, higher kurtosis suggests a heavier tail and higher peakedness at zero as $q$ increases; Figure~\ref{contour} depicts the joint density function of $\beta(v_1)$ and $\beta(v_2)$ for $q=1, q=3$ and $q=5$, and for different correlations. In this plot, we observe that the mass at zero increases with $q$ while they share the same covariance structure with unit variance term. 
Figure 1 of the Supplementary Material shows the conditional density of $\beta(v)$ for an arbitrary location $v$ given $\beta(v')$ with $v\neq v'$. The conditional density at one location tends to have a shorter peak as the value at other location moves away from zero. Also, the conditional densities tend to be more positively skewed, as we condition on higher values for the other location.

\subsubsection{Multivariate properties}
Let $P_q =Z_{1}\bigodot\ldots\bigodot Z_{q}$ be random vector of length $p$ where $Z_1\sim$ MVN$(0, \Sigma_1)$ and $Z_2,\ldots,Z_q\overset{ind}{\sim}$ MVN$(0, \Sigma_2)$ and $\Sigma_2$ has diagonal entries equal to one. Then we have $E(P_q'P_q)=\text{trace}(\Sigma_1)$ for all $q$ and $\Sigma_2\bigodot\ldots(q\text{-times})\bigodot\Sigma_2\rightarrow I_p$, as $q$ increases to infinity where $I_p$ is the $p\times p$ identity matrix. The distribution of $\beta(v)$ for a finite set of locations has the above mentioned properties as $\beta_{2}(v),\ldots\cdot \beta_{q}(v)$ have the same covariance kernel with one total variance. In general, it is difficult to explicitly calculate kurtosis in multivariate setup. However, we have following alternative result.
\begin{theorem}
The multivariate kurtosis of $P_q$ increases with $q$. 
\end{theorem}
This can be proved using the method of induction and Theorem~\ref{bvt}. The proof is in the Supplementary materials.


To summarize, for $q = 1$, the PING process is the standard GP and as $q$ increases PING has a mass near zero and tail probabilities increase. Furthermore, an appropriate rescaling of the spatial correlation parameters, can maintain the smoothness properties of the original GP. Therefore, the PING process is an attractive model for a sparse and smooth signal. However, because of the differentiability property posterior mode for PING is never zero, unlike Bayesian LASSO. {  The number of terms $q$ clearly plays a major role in the application of the PING process, and we recommend to select this tuning parameter using cross-validation.} We discuss our cross validation in detail in Section~\ref{real}.

\subsection{Error distribution and Matern correlation}
\label{error}
Next, we discuss the error process $E_i(v)$. To account for both large and small scale spatial deviations of $Y_i$ we consider the following decomposition, similar to \cite{specimage} of $E_i(v)$ ,
\begin{align}
E_i(v) = \sum_{j=1}^JZ_{j}(v)\gamma_{ij} + \epsilon_i(v) \label{error1},
\end{align}
where the first term is a linear combination of known basis functions $Z_j$'s and $\gamma_{ij}$ are unknown coefficients  for $i$-$th$ visit and $j$-$th$ basis and captures the large-scale deviation. The second term $\epsilon_i$ is intended to capture small scale deviations. We assume that $\epsilon_i$ is mean-zero GP with stationary and isotropic Matern covariance function as follows:
\begin{align}
\text{Cov}(\epsilon_i(v), \epsilon_i(v')) &=C(\theta)= \sigma^2I(v=v')+\tau^2\mathcal{M}_{\nu}\Big(\frac{\|v-v'\|}{\phi}\Big), \label{Matern}
\end{align}
where $\mathcal{M}_{\nu}(h)=\frac{2^{1-\nu}}{\Gamma(\nu)}(3h\sqrt{\nu})^{\nu}\mathcal{K}_{\nu}(3h\sqrt{\nu})$ and $\mathcal{K}$ is the modified Bessel function of the second kind. The Matern covariance has four parameters $\theta=(\sigma^2,\tau^2,\phi,\nu)$, that represent the variance of the non-spatial error (nugget), the variance of the spatial process (partial sill), the spatial range and the smoothness of the correlation function respectively. 

The large-scale spatial structure is described by $J$ random-effect covariates $\{Z_1,\ldots,Z_J\}$. Among many different choices for $Z_j$'s, we consider outer product of B-spline basis functions.{  The error in approximating the non-stationary covariance function using B-splines decreases at a rate of $J^{-\alpha/2}$ \citep{shen2015adaptive}, where $J$ is the number of B-spline basis and $\alpha$ is the regularity of the process. Thus, with large $J$, this can approximate any covariance function.} We assume the random effects are normally distributed, i.e., ${\boldsymbol{\gamma_i}}=(\gamma_{i1},\ldots,\gamma_{iJ})^{T}\sim \text{Normal}(0,\Sigma)$, where $\Sigma$ is the $J\times J$ covariance matrix. The nonstationary component of the covariance is
\begin{align}
NS(v,v') = \sum_{j=1}^J\sum_{l=1}^JZ_j(v)Z_l(v')\Sigma_{jl}. \label{ns}
\end{align}
Then the overall covariance becomes sum of \eqref{Matern} and \eqref{ns}.

\section{Extension to other image regression models}
\label{extend}
The model in the previous section is designed for image-on-scalar regression. Our sparse prior can be easily adapted for other image regressions as described below. 

\subsection{Image-on-image regression}
Consider the case of a linear image-on-image regression model (see for example \cite{vcs, image1, image2, image3, Hazra})

\begin{align}
Y_{i}(v) &= \alpha(v) + \sum_{j=1}^{p}X_{ij}(v)\beta_{j}(v) + E_{i}(v),\label{matden}
\end{align}
where $Y_{i}$ is the image response and the $X_{ij}$'s are the image predictors for subject $i$. Here $\alpha(\cdot)$ is an unknown intercept as before and $\beta_{j}(\cdot)$ are spatially varying piecewise smooth and sparse covariate effects, and $E_i(\cdot)$ is the error process. 

We put the PING prior on each of $\beta_j(\cdot)$ for sparse and smooth estimation. The selection of $q$ is done based on a cross validation technique which is discussed in Section~\ref{real}. This gives local variable selection as the subset of the covariates with beta shrunk towards zero changes with $s$. 

\subsection{Scalar-on-image regression}
Finally consider the case of a scalar-on-image regression model (see \citet{Wang, kang, Goldsmith, Li}). This model is
\begin{align}
Y_{i} = \sum_{j=1}^{n}X_{i}(v_j)\beta(v_j) + \epsilon_i,
\end{align}
where $Y_i$ is the scalar response and $X_i$ is an image with $n$ spatial locations for subject $i$. Here $\beta(\cdot)$ are spatially varying piecewise smooth and sparse covariate effect, and $\epsilon_i$ is the error which follows N($0, \sigma^2$).  We again put a PING prior on $\beta(\cdot)$ for sparse and smooth estimation and its performance is studied Section~\ref{simulation}.

\section{Computational details}
The prior on intercept ($\alpha$) as well as the components in the PING prior ($\beta_{k}$'s that comprise the PING prior) are assumed to be mean-zero GP with stationary and isotropic Matern covariance function:
\begin{align}
\text{Cov}(\alpha(v), \alpha(s')) &=C(\theta_0)= \sigma_0^2I(v=v')+\tau_0^2\mathcal{M}_{\nu_0}\Big(\frac{\|v-v'\|}{\phi_0}\Big), \\
\text{Cov}(\beta_{k}(v), \beta_{k}(v')) &=C(\theta_1)= \tau_1^2\mathcal{M}_{\nu_1}\Big(\frac{\|v-v'\|}{\phi_1}\Big). \label{MaternPGP}
\end{align}
No nugget variance is assumed for the components of PING to ensure smoothness.

\label{compute}
For small and moderate datasets, standard Markov chain Monte-Carlo (MCMC) algorithms apply to the PING model and computation is straightforward. One advantage of the PING prior is the elements of the $j$-$th$ component $\big(\beta_j(v_1),\ldots, \beta_j(v_n)\big)$ have multivariate Gaussian full conditional distribution given the other $(q-1)$ GPs, and thus Gibbs steps can be used to update the PING process parameters. For large $n$ however, these updates become slow and we use spectral methods, described in the remainder of this section.

\subsection{The model in spectral domain}
\label{modelspec1}
Similar to \cite{specimage}, we partially decorrelate the data by using the discrete Fourier transformation (DFT). Let us denote spectral representation of the processes $Y_i(v), \alpha(v), \beta(v),\beta_{k}(v),$ $ X_i(v), Z_j(v)$ and $\epsilon_i(v)$ as $\tilde{Y_i}(\omega), \tilde{\alpha}(\omega), \tilde{\beta}(\omega), \tilde{\beta_{k}}(\omega),\tilde{X}_i(\omega),$ $ \tilde{Z_j}(\omega)$ and $\tilde{\epsilon_i}(\omega)$ for frequency $\omega\in \mathcal{F}\subset\mathbb{R}^3$. Since discrete Fourier transformation (DFT) preserves linearity, the spatial model in \eqref{model} in the spectral domain can be written as
\begin{align}
&\tilde{Y_i}(\omega) = \tilde{\alpha}(\omega) + \tilde{X}_i(\omega)*\tilde{\beta}(\omega) + \sum_{j=1}^J \tilde{Z_j}(\omega)\gamma_{ij} + \tilde{E_i}(\omega),\label{modelspec}\\
&\tilde{\beta}(\omega) = \tilde{\beta}_{1}(\omega)* \tilde{\beta}_{2}(\omega)*\ldots*\tilde{\beta}_{q}(\omega).\label{PGPspec}
\end{align}
The notation $*$ denotes convolution. The Gaussian process $\alpha(v), \beta_{k}(v)$ and $\epsilon_i(v)$ are stationary and defined over a discrete spatial domain. In order to avoid computationally expensive Bessel function and spectral aliasing calculations, we use the quasi Matern spectral density \citep{Guni1}, which mimics the flexibility of the Matern spectral density for $ \tilde{\alpha}(\omega), \tilde{\beta}_{k}(\omega)$ and $\tilde{\epsilon_i}(\omega)$,
\begin{align}
\lambda(\omega|\theta=(\sigma^2,\tau^2,\phi,\nu))=\sigma^2 + \tau^2\big[\frac{1}{\phi^2}+h(\omega)\big]^{-\nu - d/2}\label{specMatern},
\end{align}
where $d$ is the dimension, $\omega\in[0,2\pi]^d$ and $h(\omega)=\sum_{j=1}^d \sin(\omega_j/2)^2$. More specifically,
\begin{align}
\tilde{E_i}(\omega)&\sim\textrm{Normal}(0,\tilde{\lambda}(\omega|\theta)),\label{specden}\\
\tilde{\alpha}(\omega)&\sim\textrm{Normal}(0,\tilde{\lambda}(\omega|\theta_0)),\nonumber\\
\tilde{\beta}_{k}(\omega)&\sim\textrm{Normal}(0,\tilde{\lambda}(\omega|\theta_1)),\nonumber
\end{align}
where $\tilde{\lambda}(\omega|\theta))={\lambda}(\omega|\theta))/2$ if $\omega\in\{0,\pi\}^3$ and $\tilde{\lambda}(\omega|\theta)={\lambda}(\omega|\theta)$ otherwise. All the parameters in $\theta$ have the same interpretation as in \eqref{Matern}. For $\tilde{\beta}_{k}$, the nugget variance is zero.

\subsection{Imputation method}
\label{impute}
Each spectral element $Y_i(\omega)$ is a function of $Y_i(v)$ for all $v \in \mathcal{V}$, and thus spectral methods require complete data. However, in practice, data are often not collected on a complete regular grid and thus the response $Y_i(v)$ is missing at many locations $v \in \mathcal{V}$ if we transform it into a regular grid. For example, in brain images we consider the complete regular grid as the 3D cube in which the skull is inscribed; from this perspective the medical images involve missingness. Missing values are handled naturally in a Bayesian context within a Gibbs sampler that draws the missing values from their conditional distribution given the observed data and the other parameters. Because imputation is applied during each MCMC iteration to account for imputation uncertainty, this step must be computationally efficient.

Denote the conditional mean by
$$
\mu_i(v) = \alpha(v) + X_i(v)\beta(v) + \sum_{j=1}^JZ_{j}(v)\gamma_{ij},
$$
and define $Y_{i1}$ to be the vector of observed data for subject $i$ and $Y_{i2}$ to be the vector representing the missing values. Likewise, let $\mu_{i1}$ and $\mu_{i2}$ be the corresponding vectors of means. The conditional distribution of $(Y_{i1}, Y_{i2})$ given all of the other parameters is 
\begin{align}
\begin{bmatrix}
{Y_{i1}}\\
{Y_{i2}}
\end{bmatrix}|\textrm{rest of the parameters} \sim \mathrm{Normal}\bigg(\begin{bmatrix}
{\mu_{i1}}\\
{\mu_{i2}}
\end{bmatrix}, \begin{bmatrix}
{\Sigma_{11}}\;\;\;{\Sigma_{12}}\\
{\Sigma_{21}}\;\;\;{\Sigma_{22}}
\end{bmatrix}\bigg)
\label{cov}
\end{align}
and thus the conditional distribution of $Y_{i2}$ given $Y_{i1}$ and the rest of the parameters is normal with mean $\mu_{i2} + \Sigma_{21}\Sigma_{11}^{-1}(Y_{i1}-\mu_{i1})$ and covariance $\Sigma_{22}-\Sigma_{21}\Sigma_{11}^{-1}\Sigma_{12}$.

For large datasets directly sampling from this distribution is infeasible. The limiting computational task in computing the conditional mean is solving a linear system with $\Sigma_{11}$. Since $\Sigma_{11}$ is symmetric and positive definite, this can be achieved with a preconditioned conjugate gradient (PCG) algorithm \citep{Golub}, an iterative method for solving the linear system $\Sigma_{11}a=b$. The goal of iterative linear solvers is to generate a sequence $a_1, a_2,\ldots$ that converges to $a=\Sigma_{11}^{-1}b$. The algorithms generally require us to compute $\Sigma_{11}a_k$ at each iteration $k$ to check for convergence and to generate the next vector in the sequence, and thus the algorithms are fast when this forward multiplication can be computed quickly. In this case, forward multiplications with $\Sigma_{11}$ can be computed in $O(n\log n)$ time and $O(n)$ memory with circulant embedding algorithms \citep{wood}, as can the forward multiplication with $\Sigma_{21}$. This is because $\Sigma_{11}$ and $\Sigma_{21}$ can be embedded in the larger circulant matrix $\Sigma$, that is,
\begin{align*}
\Sigma\begin{bmatrix}
{a_k}\\
{0}
\end{bmatrix}=\begin{bmatrix}
{\Sigma_{11}}\;\;\;{\Sigma_{12}}\\
{\Sigma_{21}}\;\;\;{\Sigma_{22}}
\end{bmatrix}\begin{bmatrix}
{a_k}\\
{0}
\end{bmatrix}= \begin{bmatrix}
\Sigma_{11}{a_k}\\
\Sigma_{21}{a_k}
\end{bmatrix},
\end{align*}
and fast Fourier transform can be exploited to compute the forward multiplication with the (nested block) circulant matrix $\Sigma$, since (nested block) circulant matrices are diagonalizable by the ($d$-dimensional) DFT. The preconditioned conjugate gradient algorithm uses an approximate inverse of $\Sigma_{11}$, called a preconditioner, to encourage the sequence $a_k$ to converge to $a$ is a small number of iterations.

Completing the imputation step requires us to simulate a residual vector with covariance matrix $\Sigma_{22}-\Sigma_{21}\Sigma_{11}^{-1}\Sigma_{12}$. To accomplish this, we first simulate a vector $(\varepsilon_{i1},\varepsilon_{i2})$ with mean zero and covariance as in \eqref{cov}, which is again efficient with circulant embedding. Then we form and the residual $\varepsilon_{i2}-\Sigma_{21}\Sigma_{11}^{-1}\varepsilon_{i1}$, which has the desired and can be computed in the same fashion as the conditioned mean. Further computation details for the conditional draws can be found in \cite{Stroud} and \cite{Guni1}.

\subsection{Sampling}
Total variances are updated from their posterior inverse gamma distributions. All other Mat\'ern parameters are updated using Metropolis sampling.  The discrete Fourier transformation (DFT) of the PING process parameters is the convolution of frequencies as in \eqref{PGPspec}. Conducting a full conditional Gibbs update, even in the spectral domain, is computationally expensive. One can improve the computational efficiency using PCG as described in Algorithm 1 of the supplementary materials. However, for our real-data application, this still imposes a serious computational burden. The existing Metropolis techniques for a joint update of large coefficient vectors, such as the gradient adjusted Metropolis-Hastings \citep{MHG} or Hamiltonian Monte Carlo \citep{HMC} mix slowly. Here we introduce a new sampling technique that uses Metropolis steps for updating each $\beta_k(\cdot)$.
The general $q$-component model can be written as
\begin{align}
Y_i(v)=\alpha(v)+\beta_{k}(v)\beta_{-k}(v)t_i+ E_i(v), \label{gibborig}
\end{align}
where $\beta_{-k}=\beta_{1}\cdots\beta_{k-1}\cdot\beta_{k+1}\cdots\beta_q$.
Denote the estimated values at the $N$-${th}$ stage of the MCMC iteration as $Y^{N}$ (samples using PCG), $\alpha^{N}$, $\beta_{k}^{N}$ and $\beta_{-k}^{N}$. We can calculate the error at $N$-${th}$ stage as $E^{N}=Y^{N}-\alpha^{N}-\beta_{k}^{N}\beta_{-k}^{N}t$.

We can rewrite our model in \eqref{gibborig} as $$\frac{Y_i(v)}{\beta_{-k}(v)}=\frac{\alpha(v)}{\beta_{-k}(v)}+\beta_{k}(v)t_i+E_i(v)\bigg(\frac{1}{\beta_{-k}(v)} - 1\bigg) + E_i(v).$$
Except for $\beta_k(v)$ and the last $E_i(v)$, replacing all other values by the ones from the $N^{th}$ step gives
\begin{align}
\frac{Y^{N}_i(v)}{\beta_{-k}^{N}(v)}=\frac{\alpha^{N}(v)}{\beta_{-k}^{N}(v)}+\beta_{k}(v)t_i+E^{N}_i(v)\bigg(\frac{1}{\beta_{-k}^{N}(v)} - 1\bigg) + E_i(v) \label{gibb}
\end{align}
The notation $Y^{N}_i$ denotes the a full response dataset including imputed missing values from the $N$th iteration. To update ${\beta}_{k}(\omega)$, we sample $\tilde{\beta}_k^u(\omega)$ according to step (ii) from Algorithm \ref{algo}, take the inverse DFT to obtain $\beta_k^u(v)$, and then form the Metropolis candidate
$\beta_{k}^{N}(v)+c\frac{\beta_{k}^{u}(v)-\beta_{k}^{N}(v)}{\|\beta_{k}^{u}(v)-\beta_{k}^{N}(v)\|_2}$. Here $c$ acts as a tuning parameter and $\|\cdot\|$ denotes the $\ell_2$ norm, defined as $\|\beta\|_2^2=\int_{v\in\mathcal{V}}\beta^2(v)dv$. We are essentially sampling $\tilde{\beta}_{k}^{u}(\omega)$ from an approximated model and then shrinking it back towards $\tilde{\beta}_k^N(\omega)$. Smaller values of $c$ generate higher acceptance rate and vice versa.
This step is described in detail in Algorithm~\ref{algo}.

\begin{algorithm}[H]
\SetAlgoLined
(i) Calculate $Q_i(v) = \frac{Y^{N}_i(v)}{\beta_{-k}^{N}(v)}-\frac{\alpha^{N}(v)}{\beta_{-k}^{N}(v)}-E^{N}_i(v)\big(\frac{1}{\beta_{-k}^{N}(v)} - 1\big)$. The superscript $N$ denotes the values at $N$-$th$ step on the MCMC. Transform $Q_i(v)$ into spectral domain to get $\tilde{Q}_i(\omega)$.\\

(ii) Generate ${\tilde{\beta}}_k^u(\omega)\sim \mathrm{Normal}(M(\omega),V(\omega))$, where $V(\omega)=\frac{1}{\sum_i(t_i^2)}(1/\tilde{\lambda}(\omega|\theta_k)+1/\tilde{\lambda}(\omega|\theta))^{-k}$, where $\tilde{\lambda}(\omega|\theta_k)$ and $\tilde{\lambda}(\omega|\theta)$ are the spectral variances of the prior on ${\tilde{\beta}}_k(\omega)$ and the error process ${\tilde{E}}_i(\omega)$ respectively as described in Section~\ref{modelspec1} with $\theta_1$ and $\theta$ as corresponding Matern parameters and $M(\omega)=\sum_i\tilde{Q}_i(\omega)/V(\omega)$.\\

(iii) Convert ${\tilde{\beta}}_k^u(\omega)$ into spatial domain ${{\beta}}_k^u(v)$ using spectral methods for spatial data as in Section 2 of \cite{reich2014spectral}.\\

(iv) Adjust the update ${{\beta}}_k^c(v)=\beta_{k}^{N}(v)+c\frac{\beta_{k}^{u}(v)-\beta_{k}^{N}(v)}{\|\beta_{k}^{u}(v)-\beta_{k}^{N}(v)\|_2}$.\\
(v) Convert $\beta^c={{\beta}}_k^c(v)\cdot\beta_{-k}^N(v)$ back into spectral domain to get ${\tilde{\beta}^c}$, again by using spectral methods for spatial data as in Section 2 of \cite{reich2014spectral}.\\
(vi) Calculate the acceptance probability for the MH step, $$P_{\beta_k^c,\beta_k^N}=\min\bigg\{1,\frac{\exp(-\sum_{i=1}^n\|(\tilde{Y}_i^N-\tilde{\beta}^ct_i)/\tilde{\lambda}(.|\theta)/\|_2^2-\|\tilde{\beta}_k^c/\tilde{\lambda}(.|\theta_k)\|_{2}^2)}{\exp(-\sum_{i=1}^n\|(\tilde{Y}_i^N-\tilde{\beta}^ct_i)/\tilde{\lambda}(.|\theta)/\|_{2}^2-\|\tilde{\beta}_k^N/\tilde{\lambda}(.|\theta_k)\|_{2}^2)}\bigg\}.$$
  \caption{Sampling algorithm of $\beta_1$ in spectral domain for the model $Y_i(v)=\alpha(v)+\beta_1(v)\beta_{-1}(v)t_i+E_i(v)$}
 \label{algo}
\end{algorithm}

In our simulation, we adjust $c$ to maintain an acceptance rate of around 0.6 for this scheme to ensure good mixing. If acceptance is lower than 0.55, we lower the value of $c$, and if it is higher than 0.70, we increase $c$ under the restriction that $c\leq \|\beta_{1}^{u}-\beta_{1}^{N}\|_2$. If we start the algorithm with small $E_i(v)$, the convergence is faster. To ensure that, we propose to get least squares estimates of $\alpha$ and $\beta$ from the data. Then assign $\beta_i=\beta^{1/q}$ as the starting value if $q$ is odd. We also recommend to take odd $q$ so that it will be easier to take $1/q$-$th$ power, although this is not necessary for starting from the least squares estimate. In the supplementary materials, Figure 7 illustrates the output of MCMC chain at 5000-$th$ iteration with a random starting point for our algorithm as well as HMC and shows that our algorithm performs overwhelmingly better than HMC and also with respect to the speed as well.

We use this spectral method for all image-on-scalar regressions in this paper.  For the simulated image-on-image and scalar-on-image regressions in Section~\ref{simulation} the datasets are small, so we use Gibbs sampling for the PING process parameters.  For larger problems, the Metropolis scheme explained above could also be adapted to image-on-image and scalar-on-image regressions.

\section{Simulation results}
\label{simulation}
In this section, we present simulation results for all three regression setups, namely image-on-scalar regression, image-on-image regression, and scalar-on-image regression. We compare the results in terms of mean squared error (MSE) with respect to the posterior mean, true positive (TP) and false positive (FN) for different levels of signal to noise ratios (SNR). However, as we use 95\% credible intervals, constructed from the 0.025 and 0.975 quantiles of the MCMC samples at each voxel separately to determine significance, this does not cause any practical issue. We calculate credible intervals of the estimates to get the proportions TP and FN. TP is defined as the proportion of locations such that the credible interval does not include zero given the true parameter value is non-zero. FN is defined as the proportion of locations such that the credible interval does not include zero given the true parameter is zero.

\subsection{Image-on-scalar regression}
Here we consider the image-on-scalar regression model in Section~\ref{model} for images of dimension $20\times 20\times 20$ with $20$ visits. The model is
\begin{align}
Y_{i}(v) = \alpha(v) + t_i\beta(v) + e_{i}(v), \label{sim1}
\end{align}
here $v \in \{1,\ldots,20\}^3$ with $i=1,2,\ldots,20$ and $t_i$'s are 20 equidistant points such that $\sum_it_i=0$ and $\sum_it_i^2=20$ obtained by standardizing the times $i=1,2,\ldots,20$. The true signal is zero for most of the spatial locations but has subregions that are non-zero. Let, $d_1=(6, 14, 6), d_2 = (6, 10, 14), d_3 = (14, 6, 14), d_4 = (14, 14, 14)$ and $d_5 = (6, 6, 6)$, $\kappa(v) = 2[\exp(-4\|v-d_1\|_2^2/20)+\exp(-1.5\|v-d_2\|_2^2/20)+\exp(-4\|v-d_3\|_2^2/20)+\exp(-4\|v-d_4\|_2^2/20)+\exp(-4\|v-d_5\|_2^2/20)$. The true signal is $\beta(v)=\kappa(v)$ if $\kappa(v)\geq 0.1$ and $\beta(v)=0$ otherwise. The plot of the true slope $\beta(v)$ is in the Supplementary Material (Figure 2). The error process $e_i(v)$ is assumed to be GP with stationary Matern covariance function. The true reparametrized Matern parameters for intercept process $\alpha(v)$ are $(1, 0.95, 10, 1)$ and last three parameters for error are $(0.90,10,1)$. After generating the data on $20\times 20\times 20$ grid, we treat the values missing outside of the inner grid $18\times 18\times 18$. Here we use the imputation technique to impute those missing values in our estimation. 

We put Gaussian process prior with Matern covariance function on the intercept process $\alpha(v)$ and PING prior (Section~\ref{PING}) on the slope. We represent $\theta =(\sigma^2,\tau^2,\phi,\nu)$, $\theta_0 =(\sigma_0^2,\tau_0^2,\phi_0,\nu_0)$ and $\theta_1 =(\sigma_1^2,\tau_1^2,\phi_1,\nu_1)$ as the Matern parameters for the error, intercept and first component in PING process prior on slope respectively. For other components of the PING prior the Matern parameters are $(1,\tau_1^2,\phi_1,\nu_1)$. We reparametrize the Matern parameter $\theta =(\sigma^2,\tau^2,\phi,\nu)$ to $\theta' = (\vartheta^2,\zeta^2,\phi,\nu)$ as $\vartheta^2 = (\sigma^2+\tau^2)$ and $\zeta^2 = \frac{\tau^2}{\sigma^2+\tau^2}$. Here $\vartheta^2$ is called the total variance. The total variance of error is set at 0.09, 0.017 and 0.009 to achieve different SNRs which are mentioned in the Table~\ref{f1}. All these results, compiled in Table~\ref{f1} are based on 50 replications and 10000 post burn samples after burning in 10000 samples. 

We fit the model with $q=1,3,5$ and priors : $\vartheta^{-2}, \vartheta_0^{-2},  \vartheta_1^{-2} \sim$ Gamma(0.1, 0.1);\\ $\text{logit}(\zeta), \log\phi, \log\nu \sim $ N(0,1) and $\text{logit}(\zeta_0),$ $ \log\phi_0, \log\nu_0 \sim $ N(0,1). For PING process: We set $\zeta_1=1$ (as nugget variance is zero) and $\log\phi_1, \log\nu_1 \sim $ N(0,1). The priors are same for the next two regressions as well.
    
From the values in the Table~\ref{f1}, we infer that for lower SNR, more components in the PING process prior leads to better estimation. Figure 3 of Supplementary Material compares the estimates for one slice of the 3-D slope across different methods along with the true $\beta(v)$. Gaussian process prior overestimates the regions where the true value is zero as shown in Figure 3 of Supplementary Material. This results in higher false positive and higher MSE for locations where the true value is zero. We calculate the MSE for $\beta=0$ as $\sum_{s:\beta_0(s)=0}(\beta(s))^2$ and total MSE as $\sum_{s}(\beta(s)-\beta_{0}(s))^2$, where $\beta_0(\cdot)$ is the truth. Here all methods have high power. {\it {In the supplementary materials we examine the effectiveness of using out-of-sample prediction to selection q.  Supplementary Figure 5 shows that prediction MSE indeed selects q=5 in this case, which is optimal.}}

\begin{table}[htbp]
    \centering
    \caption{Total MSE, MSE for the subregion with true $\beta=0$ along with standard errors in the bracket, power, coverage and false positive for the slope of the image-on-scalar simulation with different SNRs for Gaussian, PING 3 and PING 5 as choices of prior} 
    \resizebox{0.8\textwidth}{!}{\begin{minipage}{\textwidth}
            \begin{tabular}{r|r|llr}
                \hline
                \multicolumn{2}{l|}{} &\multicolumn{3}{l}{Fitted Model} \\ \hline
                SNR &Metric  & Gaussian & PING 3 &PING 5 \\ 
                \hline
                
                &Total MSE & $0.22\times10^{-2} $&$3.08\times10^{-2}$&$0.23\times10^{-2}$\\
                &&$(0.25\times10^{-3})$&($13.09\times10^{-3}$)&$(1.42\times10^{-3})$\\
                1&MSE for $\beta=0$ & $2.49\times10^{-3}$&$4.34\times10^{-5}
                $&$1.25\times10^{-3}$\\
                &&$(3.61\times10^{-4})$&($0.38\times10^{-4}
                $)&$(5.38\times10^{-4})$\\
                &True positive  &0.97&0.94&0.94\\
                &False positive & $0.63\times 10^{-1}$&$0.01\times 10^{-1}$&0.00\\
                 &Coverage &0.92&0.70&0.98\\
                \hline
                
                &Total MSE & $9.63\times10^{-4} $&$12.07\times10^{-4}$&$4.19\times10^{-4}$\\
                &&$(1.19\times10^{-4})$&($11.44\times10^{-4}$)&$(0.49\times10^{-4}$\\
                5&MSE for $\beta=0$ & $14.95\times10^{-4}$&$3.84\times10^{-4}$&$4.10\times10^{-4}$\\
                &&$(2.35\times10^{-4})$&($1.87\times10^{-4}$)&$(0.56\times10^{-4})$\\
                &True positive  &1&1&1\\
                &False positive & $2.19\times 10^{-1}$&0.00&0.00\\
                &Coverage &0.88&0.98&0.99\\
                \hline
                
                &Total MSE & $7.58\times10^{-4} $&$3.79\times10^{-4}$&$2.25\times10^{-4}$\\
                &&$(8.17\times10^{-5})$&($3.09\times10^{-5}$)&$(2.64\times10^{-5}$\\
                10&MSE for $\beta=0$ & $11.94\times10^{-4}$&$2.49\times10^{-4}$&$2.25\times10^{-4}$\\
                &&$(2.11\times10^{-4})$&($9.57\times10^{-5}$)&$(3.06\times10^{-5}$\\
                &True positive  &1&1&1\\
                &False positive & $2.72\times 10^{-1}$&0.00&0.00\\ &Coverage &0.85&0.98&0.99\\
                
                \hline
            \end{tabular}
    \end{minipage}}
    \label{f1}
\end{table}

\subsection{Image-on-image regression}
We consider image-on-image regression model as in Section~\ref{extend} 
\begin{align}
Y_{i}(v) = \alpha(v) + \sum_{j=1}^{10}X_{ij}(v)\eta_{j}(v) + e_{i}(v), \label{sim2}
\end{align}
on data collected over 100 locations, selected at random in $[0,1]^2$ with $i=1,\ldots,20$ observations at each location. The ten spatially varying predictors ($X$'s) are generated using the reparametrized Matern parameters, generated randomly. First a random vector of four elements are generated from N(0, 1). We exponentiate first, third and fourth element and take inverse logit transformation of the second element to get those reparametrized Matern parameters for each predictor. These predictors are generated only once for the whole simulation. We change the domain of the images for this simulation from previous case and only consider data at 100 locations to construct a dataset of manageable dimension for easier computation.

The error process $e_i(v)$ is assumed to be GP with stationary Matern covariance function, independent over $i$. And $\eta_{j}(v)=0$ for $j=1,\ldots, 5$. Rest of those five $\eta$'s have the structures, plotted in Figure 4 of the Supplementary Material. These are zero at most of the locations with some non-zero subregions. To generate $\eta_6$, we divide the whole $[0, 1]^2$ space into a $50\times 50$ grid. Then we generate a random number $h$ in $\{1,2,3\}$. Then we generate $h$ set of co-ordinates in $[0, 1]^2$. Let these be $u_1,\ldots,u_h$. Let us define $\kappa(v)=\sum_{i=1}^h2\exp(-3\|v-50u_i\|_2^2/50)$. Then $\eta_6(v)=\kappa(v)$ if $\kappa(v)\geq 0.1$ and $\eta_6(v)=0$ otherwise. Other four $\beta_j$'s are generated similarly. The true reparametrized Matern parameters for intercept are $(1,0.95,10,1)$ and last three parameters for error are $(0.9,10,1)$. The total variance of error is set to 0.57, 0.11 and 0.06 to achieve different SNRs which are mentioned in the table. We report MSE, power, false positive and coverage averaged over $\eta$. All these results, compiled in Table~\ref{f2} are based on 50 replications and 5000 post burn samples after burning in 5000 samples.

In Table~\ref{f2}, we see that the PING process prior always gives a better estimate in terms of MSE. The GP prior overestimates the regions where the true value is zero. This results in higher false positive for GP prior. Here all methods have similar power. As the SNR increases, the results using PING are even better than those using Gaussian.

\begin{table}[htbp]
    \centering
    \caption{Total MSE, MSE for the subregion with true $\eta=0$ along with standard errors in the bracket, power, coverage and false positive error for the slope of the image-on-image simulation with different SNRs for Gaussian, PING 3 and PING 5 as choices of prior} 
    \resizebox{0.8\textwidth}{!}{\begin{minipage}{\textwidth}
            \begin{tabular}{r|r|llr}
                \hline
                \multicolumn{2}{l|}{} &\multicolumn{3}{l}{Fitted Model} \\ \hline
                SNR &Metric  & Gaussian & PING 3 compo & PING 5 compo \\ 
                \hline
                
                &Total MSE & $3.69\times 10^{-3}$&$1.14\times 10^{-3}$&$0.97\times 10^{-3}$\\
                &&$(0.44\times 10^{-3})$&($0.22\times 10^{-3}$)&$(0.19\times 10^{-3})$\\
                1&MSE for $\eta=0$ & $2.91\times 10^{-3}$&$0.67\times 10^{-3}$&$0.49\times 10^{-3}$\\
                &&$(4.25\times 10^{-4})$&($1.63\times 10^{-4}$)&$(1.37\times 10^{-4})$\\
                &True positive  &0.87&0.86&0.84\\
                &False positive &$0.20\times 10^{-1}$&$0.02\times 10^{-1}$&$0.01\times 10^{-1}$\\
                &Coverage &0.97&0.99&0.99\\
                \hline  
                &Total MSE & $9.71\times 10^{-4}$&$ 2.56\times 10^{-4} $&$2.04\times 10^{-4}$\\
                &&($1.19\times 10^{-4}$)&($0.51\times 10^{-4}$)&($0.46\times 10^{-4}$)\\
                5&MSE for $\eta=0$ & $8.82\times 10^{-4}$&$3.38\times 10^{-4}$&$1.22\times 10^{-4}$\\
                &&($11\times 10^{-5}$)&($8.51\times 10^{-5}$)&($3.14\times 10^{-5}$)\\
                &True positive  &0.97&0.96&0.96\\
                &False positive &$0.29\times 10^{-1}$&$0.02\times 10^{-1}$&$0.01\times 10^{-1}$\\
                &Coverage &0.97&0.99&0.99\\
                \hline
                & Total MSE & $5.36\times 10^{-4}$&$1.36\times 10^{-4}$&$1.06\times 10^{-4}$\\
                &&($6.32\times 10^{-5}$)&($2.65\times 10^{-5}$)&($2.50\times 10^{-5}$)\\
                10 &MSE for $\eta=0$ & $5.00\times 10^{-4}$&$0.98\times 10^{-4}$&$0.66\times 10^{-4}$\\
                &&($5.88\times 10^{-5}$)&($1.96\times 10^{-5}$)&($1.70\times 10^{-5}$)\\
                &True positive  &0.98&0.98&0.97\\
                &False positive &$0.33\times 10^{-1}$&$0.03\times 10^{-1}$&$0.01\times 10^{-1}$\\
                &Coverage &0.96&0.99&0.99\\
                \hline
            \end{tabular}
    \end{minipage}}
    \label{f2}
\end{table}

\subsection{Scalar-on-image regression}

Finally we replicate the simulation from \citep{kang} with 100 observations.  For each observation, there is a two-dimensional image $X_i$ of dimension $20\times 20$ with an exponential covariance structure having range parameter 3. The model is
\begin{align}
Y_{i} \overset{ind}{\sim}\text{Normal}(\sum_{j,k=1}^{20}X_{ijk}\beta_{jk}, \sigma^2), \label{sim3}
\end{align}
Here the coefficient $\beta=(\!(\beta_{jk})\!)_{1\leq j,k\leq 20}$ is a matrix of dimension $20\times 20$. The true $\beta$ is generated in such a way that it has five peaks. Let, $d_1 = (4, 16), d_2=(16, 4), d_3=(4,4), d_4=(16, 16)$ and $d_5=(10, 10)$ and $\kappa(v) = \sum_{1}^{5}2\exp(-20\|v-d_i\|_2^2/50)$. Then mathematically the true beta is $\beta(v)=\kappa(v)$ if $\kappa(v)\geq 0.1$ and $\beta(v)=0$ otherwise. Only in this set up we have number of observations much less than number of parameters to be estimated. We consider three choices of $\sigma^2$ in generating the data, 0.1, 1 and 1.5.

\begin{figure}[htbp]
    \centering
    \includegraphics[width = 0.6\textwidth]{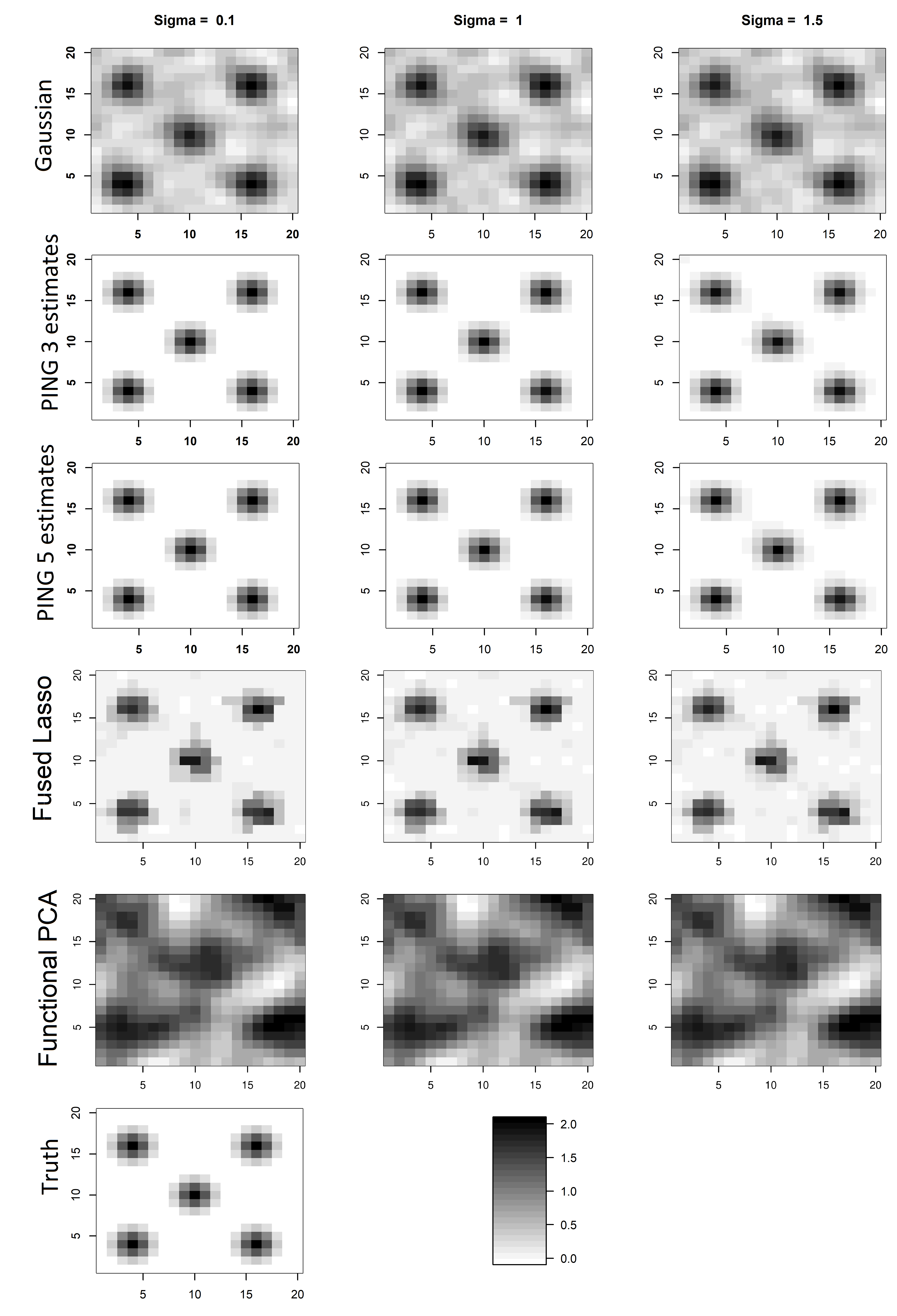}
    \caption{Comparison plot of the estimates for the slope of the scalar-on-image simulation with different true variances for Gaussian, PING 3 and PING 5 as choices of prior along with fused lasso and functional PCA estimates.}
    \label{heatmap3rdsim3}
\end{figure}

We fit the model with $q=1,3,5$. The prior for $\sigma^{-2}$ is $\mathrm{Gamma}(0.1, 0.1)$. Rest of parameters have the same prior as in previous subsections. {  In this subsection we also compare our method with fused lasso \citep{fused} and functional principal component analysis (fPCA) \citep{jones1992displaying}. Fused lasso estimates are computed using an archived version of previously available {\tt genlasso} package in R. After smoothing the images using {\tt fbps} function of {\tt refund} package, eigendecomposition of the sample covariance is computed. After that lasso regularized principal components regression is performed. The leading eigenvectors that explain 95\% of the variation in
the sample images are used to get the final estimate.

We report MSE, true positive, false positive and coverage in the estimation of the slope $\beta$ matrix. All these results, compiled in Table~\ref{f3}, are based on 50 replications and 5000 post-burn samples after burning in 5000 samples.
 Here, we see that the estimates from the PING process prior are superior to those of the GP, fused lasso and functional PCA (fPCA) in all metrics. In particular, fPCA results are very noisy. In Figure~\ref{heatmap3rdsim3} the estimated parameters from PING are less noisy than all other estimates.

We also compare PING with STGP method \citep{kang}. While comparing with STGP, we consider the low-rank approximation of each component of PING. The low-rank approximation is incorporated following the works in the kernel convolution of \cite{higdon} and \cite{nychka} as in \cite{kang}. Due to this modification, all the results change from the previous table and the comparison results with STGP are provided separately in Table~\ref{f31}. We only show results using PING-8, as they are the best based on MSEs in a grid of choices, ranging from 2 to 20. In the supplementary materials, a plot of MSEs for different choices of $q$ is given.

We can see that the estimates from PING process prior are far better than STGP in terms of overall MSE, true positive and coverage except for the MSE at the subregion where the truth is zero. The STGP and PING estimates are almost indistinguishable in the picture. Due to thresholding, STGP estimates are more conservative. Thus, a pictorial comparison will not be much informative and we have not presented it.}

\begin{table}[htbp]
    \centering
    \caption{Total MSE, MSE for the subregion with true $\beta=0$ along with standard errors in the bracket, power, coverage and false-positive error for the slope of the scalar-on-image regression model with different true variances for Gaussian, PING 3 and PING 5 as choices of prior along with fused lasso and functional PCA estimates}
    \resizebox{0.75\textwidth}{!}{\begin{minipage}{\textwidth}
            \begin{tabular}{r|r|llrrr}
                \hline
                \multicolumn{2}{l|}{} &\multicolumn{5}{l}{Fitted Model} \\ \hline
                $\sigma$ &Metric& Gaussian & PING 3 & PING 5 & Fused lasso & Functional PCA\\ 
                \hline 
                & Total MSE & $3.01\times 10^{-2}$ & $0.62\times 10^{-2}$ & $0.88\times 10^{-2}$ &$2.82\times 10^{-2}$&7605.560\\ 
                & & $(1.8\times 10^{-3})$ & $(2.1\times 10^{-3})$ & $(2.8\times 10^{-3})$&$(76.9\times 10^{-3})$& (7680.645)\\ 
                &MSE for $\beta=0$ & $2.01\times 10^{-2}$ & $0.21\times 10^{-2}$ & $0.19\times 10^{-2}$ &$0.70\times 10^{-2}$&6495.84\\ 
                & & $(1.32\times 10^{-3})$ & $(0.66\times 10^{-3})$ & $(0.82\times 10^{-3})$ &$(29.6\times 10^{-3})$&7358.84\\ 
                1.5&True positive & $0.79$ & $0.75$ & $0.68$&0.56&1 \\ 
                &False positive & $8.14\times 10^{-3}$ & $0.34\times 10^{-3}$ & $0.34\times 10^{-3}$ &0.00&0.95\\ 
                &Coverage & $0.97$ & $0.96$ & $0.95$&-&- \\ 
                \hline
                &Total MSE & $2.56\times 10^{-2}$ & $0.33\times 10^{-2}$ & $0.49\times 10^{-2}$ &$2.09\times 10^{-2}$&7603.84\\ 
                & & $(1.71\times 10^{-3})$ & $(0.77\times 10^{-3})$ & $(1.51\times 10^{-3})$ &$(56.32\times 10^{-3})$&(7672.702)\\ 
                &MSE for $\beta=0$ & $1.66\times 10^{-3}$ & $0.11\times 10^{-3}$ &$ 0.09\times 10^{-3}$ &$4.36\times 10^{-3}$&6494.85\\ 
                & & $(1.12\times 10^{-3})$ & $(0.32\times 10^{-3})$ & $(0.36\times 10^{-3})$ &$(18.64\times 10^{-3})$&(7354.22)\\ 
                1&True positive & $0.86$ & $0.91$ & $0.86$ &0.69&1 \\ 
                &False positive & $14.24\times 10^{-3}$ & $0.68\times 10^{-3}$ & $1.02\times 10^{-3}$ &$3.39\times 10^{-3}$&0.96\\ 
                &Coverage & $0.96$ & $0.96$ & $0.96$ &-&-\\
                \hline
                &Total MSE & $20.90\times 10^{-3}$ & $0.31\times 10^{-3}$ & $0.50\times 10^{-3}$ &$14.73\times 10^{-3}$&7603.69\\ 
                & & $(9.79\times 10^{-4})$ & $(0.17\times 10^{-4})$ & $(0.43\times 10^{-4})$ &$(399.2\times 10^{-4})$&(7664.746)\\ 
                &MSE for $\beta=0$ & $12.75\times 10^{-3}$ & $0.18\times 10^{-3}$ & $0.20\times 10^{-3}$ &$2.28\times 10^{-3}$ &6496.13\\ 
                && $(4.38\times 10^{-4})$ & $(0.14\times 10^{-4})$ & $(0.18\times 10^{-4})$ &$(109.97\times 10^{-4})$&(7351.93)\\ 
                0.1&True positive & $0.94$ & $1.00$ & $1.00 $&0.90&1\\ 
                &False positive & $1.76\times 10^{-2}$ & $0.37\times 10^{-2}$ & $0.44\times 10^{-2}$ &0.15&1\\ 
                &Coverage & $0.94$ & $0.96$ & $0.96$ &-&-\\
                \hline
            \end{tabular}
    \end{minipage}}
    \label{f3}
\end{table}

\begin{table}
    \centering
    \caption{Total MSE, MSE for the subregion with true $\beta=0$ along with standard errors in the bracket, power and Type1 error for the slope of the scaler-on-image simulation with different true variances for soft-thresholded Gaussian process (STGP) and PING 8 as choices of prior.}
    \resizebox{0.75\textwidth}{!}{\begin{minipage}{\textwidth}
            \begin{tabular}{r|r|lr}
                \hline
                \multicolumn{2}{l|}{} &\multicolumn{2}{l}{Fitted Model} \\ \hline
                $\sigma$&Metric & STGP & PING 8 components \\ 
                \hline
                & Total MSE & $4.22\times 10^{-3}$ & $1.303\times 10^{-3}$ \\ 
                &  & ($1.57\times 10^{-3}$) & ($0.59\times 10^{-3}$) \\ 
                & MSE for $\beta=0$ & $1.04\times 10^{-4}$ & $4.82\times 10^{-4}$ \\ 
                &  & ($1.49\times 10^{-4}$) & ($1.82\times 10^{-4}$) \\ 
                1.5& True positive & 0.85 & 0.95 \\ 
                & False positive & 0.10 & 0.06 \\ 
                \hline
                & Total MSE & $3.8\times 10^{-3}$ & $0.68\times 10^{-3}$ \\ 
                &  & ($1.54\times 10^{-3}$) & ($0.28\times 10^{-3}$) \\ 
                & MSE for $\beta=0$ & $0.79\times 10^{-4}$ & $3.15\times 10^{-4}$ \\ 
                &  & ($1.28\times 10^{-4}$) & ($0.92\times 10^{-4}$) \\ 
                1& True positive & 0.86 & 0.96 \\ 
                & False positive & 0.10 & 0.06 \\ 
                \hline
                &Total MSE & $2.86\times 10^{-3}$ & $0.262\times 10^{-4}$ \\ 
                &  & ($12.59\times 10^{-4}$) & ($0.84\times 10^{-4}$) \\ 
                &MSE for $\beta=0$ & $0.27\times 10^{-4}$ & $1.17\times 10^{-4}$ \\ 
                &  & ($0.86\times 10^{-4}$) & ($0.20\times 10^{-4}$) \\ 
                0.1& True positive & 0.86 & 0.96 \\ 
                & False positive & 0.11 & 0.07 \\ 
                \hline
            \end{tabular}
    \end{minipage}}
    \label{f31}
\end{table}

In all the simulations, PING estimates work well as a shrinkage estimate, producing much better results than other alternatives.

\section{Application to longitudinal MRI data}
\label{real}

Next, we turn to the study of multiple sclerosis (MS) using MRI images. In a natural history cohort followed at the National Institute for Neurological Disorders and Stroke, each subject was scanned approximately once per month over several hospital visits. In the subset of the study published in \citep{sweeney}, several individuals were scanned over 3 years. We focus on the set of images from a single subject. Using a 1.5T GE scanner with clinically optimized scanning parameters, whole-brain magnetization transfer fluid attenuation inversion recovery (FLAIR) volumes were acquired. All the modalities were interpolated to a voxel size
of $1$ mm$^3$ yielding images of dimension $182 \times 218 \times 182$. We use normalized FLAIR images in our study by z-scoring using normal-appearing white matter \citep{shi2,shi1}. We also use Subtraction-Based Logistic Inference for Modeling and Estimation (SuBLIME) mask. The SuBLIME mask is a 4D image with three dimensions for space and one for time. For each time point, the 3D image is a map of where there were new/enlarging lesions between the corresponding pair of time points. All images were registered longitudinally and across the modalities and rigidly aligned to the Montreal Neurological Institute standard space \citep{Fonov}. \cite{sweeney} has a complete description of the study along with the acquisition parameters. Our preliminary investigation seems to indicate that the image intensity varies linearly with time. Let $Y_i(v)$ denotes the image intensity at a 3-dimensional voxel $v$ of $i$-$th$ image at time $t_i$, which denotes the number of days passed between $i$-$th$ and the first visit of a single subject. In general, $v$ is used to denote voxel. We normalize the time covariate $t_i$ and set the image of the first visit as the baseline. We consider the following model from Section~\ref{model},
$
Y_i(v) = \alpha(v) + \beta(v)t_i+ E_i(v)
$
where $\alpha(v)$ is the spatially varying intensity image at baseline visit and $\beta(v)$ quantifies the brain regions that are deteriorated over time due to MS. We decide to use this linear model in time after performing some exploratory analysis on model selection among the higher order polynomials in $t$ at each voxel. The linear model turns out to be the one with smallest AIC and BIC values. It is expected that the healthy brain tissue does not change much, while changes occur in the MS-affected brain regions and the number of such regions is small. Thus the effect $\beta(v)$ is expected to be sparse, and in addition, it is also desired to be piecewise smooth and continuous, due to the complex spatial dependence in the brain. {\it However, we have also fitted following non-linear model,
\begin{align}
   Y_i(v) = \sum_{j=1}^{K} \beta_{j}(v)B_j(t_i)+ E_i(v)\label{nonlin},
\end{align}
here $B_j(\cdot)$ stands for the B-Spline basis functions and $\beta_j(v)$ is the spatially varying coeffients. Since $\beta_1(v)$ corresponds to the effect at $0$-$th$ time point, we put PING$(1)$ which is standard Gaussian prior on $\beta_1(v)$ and PING$(q)$ on $\beta_{j}(v)$ for $j\geq 2$ and $q=3,5$. We consider linear B-spline bases with four equidistant knots after performing prediction MSE based cross validation on the degree and number of bases. } 

For the error process, we consider the non-stationary covariance model as discussed in Section~\ref{error}. We use GP prior with Matern covariance function for $\alpha$. Further details of the model specifications are: $\boldsymbol{\gamma_{i}}\sim N(0,\Sigma)$ where $\Sigma\sim IW(J+0.1, \frac{c}{J+0.1}I_{J})$ and $c^{-1}\sim$ Gamma(0.1,0.1), where ``$IW$'' stands for inverse Wishart, $J=6^3=216$ B-spline basis functions. We reduce the dimensionality of the images to $91\times 109 \times 91$ using {\tt resize} function of {\tt imager} package of R due to computational and storage issues; the reduced images preserve the overall structure of the original images. The time of the visits is roughly every month. We normalize the time vector such that the sum of squares of the times is one. We present the analysis for one MS subject in the study. Preliminary analysis that confirms linearity in the change overtime is included in the supplementary material.
The real data plot of an axial view for the subject's first 12 visits is in Figure 6 of the Supplementary Material. 

\begin{figure}[htbp]
\centering
    \includegraphics[width = 0.9\textwidth]{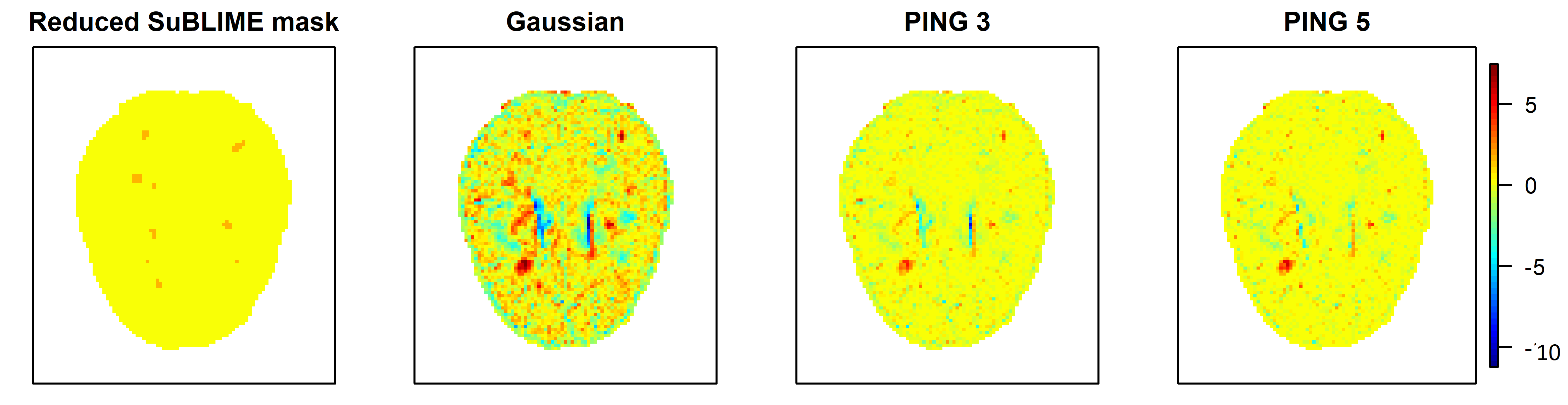}
    \caption{Estimated slope $\beta(v)$ of the middle slice using different priors along with the color scale along the reduced SuBLIME mask.}
    \label{esti59}
\end{figure}

\begin{figure}[htbp]
\centering
    \includegraphics[width = 0.5\textwidth]{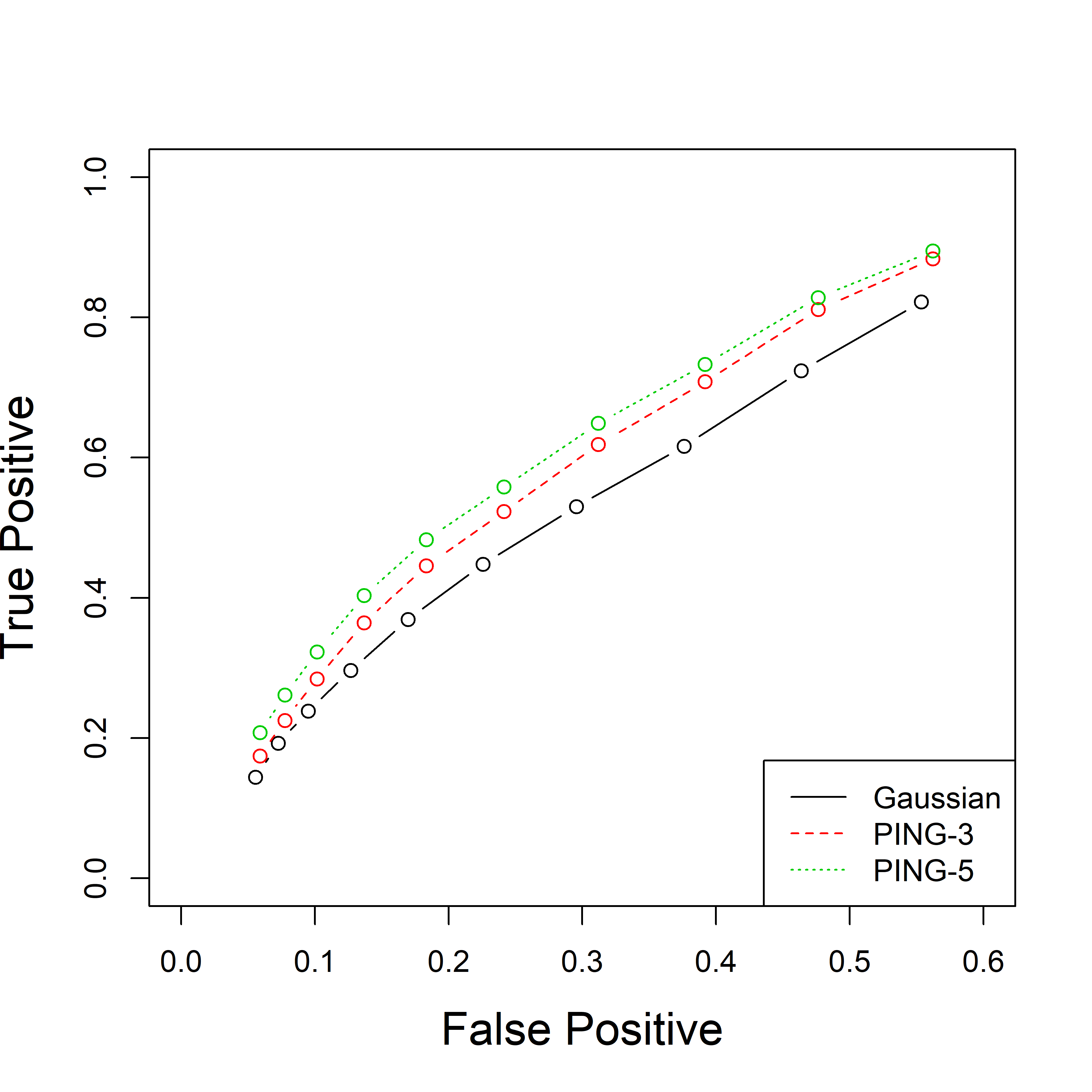}
    \caption{ROC curve constructed from the Gaussian, PING-3 and PING-5 estimates in detecting lesions flagged by reduced SubLIME mask.}
    \label{ROC}
\end{figure}

We use both the proposed method with the signal modeled using the PING process, and with the signal modeled via a GP. We consider both PING with three and five components and Gaussian prior for the slope $\beta$ and compare the estimates in terms of prediction MSE. {\it We select optimal $q$ based on a leave-one-out cross validation method. For cross validation, we consider 12 pairs of training and testing sets $\{(D_{-i}, D_i): i=1,\ldots,12\}$, where $D_{-i}=\{Y_1,\ldots,Y_{i-1},Y_{i+1},\ldots,Y_{12}\}$, $D_i=\{Y_i\}$ and $Y_i$ denotes MRI data from $i$-$th$ time point. Now, we fit the model 12 times for $D_{-i}$'s and calculate the prediction MSE $P_i$ for $D_i$. Our selection of $q$ is based on $\bar{P}$, where $\bar{P}=\frac{1}{n}\sum_i P_i$. The estimates are based on 5000 post-burn MCMC samples after 5000 burn-in. We sample the values in the image outside of the brain using the techniques of Section~\ref{impute} after every 30 iterations. There is improvement in prediction MSE using PING-5 over PING-3 and Gaussian. For the two cases PING-3 and Gaussian $\bar{P}$'s are around 0.16. For PING-5 $\bar{P}$ becomes 0.13. After PING-5, the prediction MSE deteriorates. For the non-linear model in~\eqref{nonlin}, the prediction MSE is the best for PING(3) prior on the B-spline coefficients $\beta_{j}(v)$ for $j\leq 2$.  For PING-3, $\bar{P}$ is around 0.09 which is slightly better than the $\bar{P}$ values from the linear model. For, PING-5 and Gaussian, $\bar{P}$ values are 0.12 and 0.10 respectively. The overall prediction MSE is marginally improved for the non-linear model over the linear model.} Figure~\ref{esti59} compares the estimates from Gaussian, PING-3, and PING-5 for the middle cross-section along with the reduced SuBLIME mask. The reduced SuBLIME is a 3-D mask which aggregates the original SuBLIME masks over time. Thus, this reduced SuBLIME mask identifies all the brain regions that turn into lesions between the subject's first visit to the eleventh visit. The estimated effect with a Gaussian prior is very noisy, and it is difficult to be used by practitioners who wish to identify regions affected by the MS. In contrast, the PING based estimates clearly highlight the regions of interest that are affected by the disease. The results agree with the ones obtained with the SuBLIME. {  To further investigate this, we plot the receiver operating characteristic (ROC) curve in detecting the lesions flagged by SuBLIME mask in Figure~\ref{ROC} based on the estimates of $\beta$ from the model in~\ref{modelIS}. We see that PING-3 estimate is better than Gaussian and PING-5 is slightly better than PING-3.} Due to storage issues of the posterior samples, these ROC curves are constructed based on the same level of cutoffs on the posterior estimates in detecting new lesions, flagged by the SuBLIME mask. PING estimates perform much better than the corresponding Gaussian estimate based on the ROC curve. For the non-linear model, it is difficult to prepare such plots combining the B-spline coefficients. 

\section{Conclusion and discussion}
\label{conclude}
We propose a new class of prior, entitled the PING prior, for estimating spatially sparse and smooth signals. We analyze the performance of our prior in different kinds of image regressions, namely image-on-scalar, image-on-image and scalar-on-image. We develop techniques to tackle huge dimensional datasets by transforming into the spectral domain. Our simulations show that this new prior outperforms the Gaussian prior for all the image regressions, we considered. An R package to fit PING for different image regression is available in {\url{https://github.com/royarkaprava/PING}}.

Our simulation results suggest that PING priors give better estimates than Gaussian at the locations where true value is zero. This results in lower false positive for PING. All of the methods have high true positive for both image-on-scalar and image-on-image regression models. For the scalar-on-image model, they even have better true positive than Gaussian along with lower false-positive and MSEs. The versatility in application of this prior is well studied in the simulation section of this paper. In the MRI data acquired longitudinally in a patient with MS, although there is little improvement in prediction MSE from Gaussian to PING, the disease affected areas are more easily distinguishable in PING estimates due to the shrinkage. For the non-linear model, wavelet bases can also be considered instead of B-splines \citep{torabi2007multiple, hackmack2012multi, zhang2016comparison, wang2016multiple}. In this case, selecting the PING prior for each basis coefficient allows for abrupt changes in space and time.

{\it An area of future work is the selection strategy for $q$, the number of components in PING. In this paper, our selection criterion is based on leave-one-out cross validation performance. The selected $q$ has also shown the best performance in Figure~\ref{ROC} in detecting the affected brain regions. However, one possible future direction is to put a prior on $q$ and develop computations using reversible jump MCMC (RJMCMC). RJMCMC algorithms are in general computationally expensive. To circumvent the computational complexities in RJMCMC, another possible alternative is to write the PING prior as, $$\beta(v)=\sigma\prod_{i=1}^Q\beta_i^{Z_i}(v), \quad \beta_{i}(v)\sim \textrm{GP}(0, K),\quad Z_i\sim\textrm{Ber}(p_i), \quad p_i\sim \textrm{Beta}(\alpha,\beta),$$ for some fixed large enough $Q$. Due to the huge dimensionality of MRI data, the convergence is rather slow using this prior. However, for smaller dimensional datasets, this is a possible alternative to selecting the number of components in the PING prior.}

The applicability of our proposed PING prior is not restricted to the models, described in this paper. This prior can be used to estimate any sparse and piece-wise smooth function. As long as the dimension of the data is manageable, one can develop efficient Gibbs sampler to estimate each component of the PING prior. One can also consider to use PCG to update each component of PING from its full conditional using the methods developed in Section~\ref{impute} for moderately large images exploiting conditional distribution of multivariate normal. To do that, we need to stack all the observations and one of the PING components in one large vector and get a conditional distribution of that large vector given other components of PING. Based on that expression, one can sample from the full conditional distribution of that PING component given the stack of observations using PCG. Details of this computation technique are in the supplementary materials. 

\section*{Acknowledgments}
This research was supported partially by the National Institutes of Health grant R01 ES027892 (Guinness), R01 NS085211 (Staicu), R01 MH086633 (Staicu), R21NS093349 (Shinohara), R01NS085211 (Shinohara) and R01MH112847 (Shinohara), and the
National Science Foundation grant DMS 1613219 (Guinness) and DMS 1454942 (Staicu). Shinohara was also partially supported by RG-1707-28586 from the National Multiple Sclerosis Society. The content is solely the responsibility of the authors and does not necessarily represent the official views of the funding agencies. 

We would like to thank the editor, associate editor, and the anonymous referees for their constructive comments which helped to improve the overall presentation of the manuscript.

\bigskip
\begin{center}
{\large\bf SUPPLEMENTARY MATERIAL}
\end{center}

\begin{description}

\item[Title: Supplementary materials of Sparse and smooth signal] Proof of the theorem, Tables, and Figures, referenced in Sections~\ref{model}, ~\ref{simulation} and ~\ref{real} are available in this pdf. (pdf)

\item[R-code] In submissioncode.zip file all the necessary R-codes are placed.

\item[Source-code] In the spatial-shrinkage.zip, the .tex file is kept.

\end{description}

%
%

%

\bibliographystyle{bibstyle} 
\bibliography{ssd}

\begin{thebibliography}{}

\bibitem[\protect\citeauthoryear{Armagan, Dunson, and Lee}{Armagan
  et~al.}{2013}]{Armagan}
A.~Armagan, D.~B. Dunson, and J.~Lee.
\newblock Generalized double Pareto shrinkage
\newblock \emph{Statistica Sinica}
\newblock 23 (2013): 119--143.

\bibitem[\protect\citeauthoryear{Bhattacharya, Pati, Pillai, and
  Dunson}{Bhattacharya et~al.}{2015}]{Bhattacharya}
A.~Bhattacharya, D.~Pati, N.~S. Pillai, and D.~B. Dunson.
\newblock Dirichlet-Laplace priors for optimal shrinkage
\newblock \emph{Journal of the American Statistical Association}
\newblock 110 (2015): 1479--1490.

\bibitem[\protect\citeauthoryear{Boehm-Vock, Reich, Fuentes, and
  Dominici}{Boehm-Vock et~al.}{2015}]{Boehm-Vock}
L.~Boehm-Vock, B.~Reich, M.~Fuentes, and F.~Dominici.
\newblock Spatial Variable Selection Methods for Investigating Acute Health
  Effects of Fine Particulate Matter Components
\newblock \emph{Biometrics}
\newblock 71 (2015): 167–177.

\bibitem[\protect\citeauthoryear{Carvalho, Polson, and Scott}{Carvalho
  et~al.}{2010}]{Carvalho}
C.~M. Carvalho, N.~G. Polson, and J.~G. Scott.
\newblock The Horseshoe estimator for sparse signals
\newblock \emph{Biometrika}
\newblock 97 (2010): 465--480.

\bibitem[\protect\citeauthoryear{Chen, Wang, Kong, and Zhu}{Chen
  et~al.}{2016}]{chen}
Y.~Chen, X.~Wang, L.~Kong, and H.~Zhu.
\newblock Local Region Sparse Learning for Image-on-Scalar Regression
\newblock \emph{arXiv:1605.08501}
\newblock  (2016).

\bibitem[\protect\citeauthoryear{Duane, Kennedy, Pendleton, and Roweth}{Duane
  et~al.}{1987}]{HMC}
S.~Duane, A.~D. Kennedy, B.~J. Pendleton, and D.~Roweth.
\newblock Hybrid monte carlo
\newblock \emph{Physics letters B}
\newblock 195 (1987): 216--222.

\bibitem[\protect\citeauthoryear{Fonov, Evans, McKinstry, Almli, and
  Collins}{Fonov et~al.}{2009}]{Fonov}
V.~Fonov, A.~Evans, R.~McKinstry, C.~Almli, and D.~Collins.
\newblock Unbiased nonlinear average ageappropriate brain templates from birth
  to adulthood
\newblock \emph{NeuroImage}
\newblock 47 (2009): S102.

\bibitem[\protect\citeauthoryear{Gaunt}{Gaunt}{2018}]{gaunt2018products}
R.~E. Gaunt.
\newblock Products of normal, beta and gamma random variables: Stein operators
  and distributional theory
\newblock \emph{Brazilian Journal of Probability and Statistics}
\newblock 32 (2018): 437--466.

\bibitem[\protect\citeauthoryear{Gelfand, Kim, Sirmans, and Banerjee}{Gelfand
  et~al.}{2003}]{vcs}
A.~Gelfand, H.~Kim, C.~F. Sirmans, and S.~Banerjee.
\newblock Spatial Modeling With Spatially Varying Coefficient Processes
\newblock \emph{Journal of the American Statistical Association}
\newblock 98 (2003): 387--396.

\bibitem[\protect\citeauthoryear{Goldsmith, Huang, and Crainiceanu}{Goldsmith
  et~al.}{2014}]{Goldsmith}
J.~Goldsmith, L.~Huang, and C.~M. Crainiceanu.
\newblock Smooth scalar-on-image regression via spatial Bayesian variable
  selection
\newblock \emph{Journal of Computational and Graphical Statistics}
\newblock 23 (2014): 46--64.

\bibitem[\protect\citeauthoryear{Golub and Van~Loan}{Golub and
  Van~Loan}{2012}]{Golub}
G.~H. Golub and C.~Van~Loan.
\newblock Matrix computations
\newblock \emph{JHU press}
\newblock  (2012): chapter 10.

\bibitem[\protect\citeauthoryear{Griffin and Brown}{Griffin and
  Brown}{2010}]{Griffin}
J.~E. Griffin and P.~J. Brown.
\newblock Inference with Normal-Gamma prior distributions in regression
  problems
\newblock \emph{Bayesian Analysis}
\newblock 5 (2010): 171--188.

\bibitem[\protect\citeauthoryear{Guinness and Fuentes}{Guinness and
  Fuentes}{2017}]{Guni1}
J.~Guinness and M.~Fuentes.
\newblock Circulant embedding of approximate covariances for inference from
  Gaussian data on large lattices
\newblock \emph{Journal of Computational and Graphical Statistics}
\newblock 26 (2017): 88--97.

\bibitem[\protect\citeauthoryear{Hackmack, Paul, Weygandt, Allefeld, Haynes,
  Initiative, et~al.}{Hackmack et~al.}{2012}]{hackmack2012multi}
K.~Hackmack, F.~Paul, M.~Weygandt, C.~Allefeld, J.-D. Haynes, A.~D.~N.
  Initiative, et~al.
\newblock Multi-scale classification of disease using structural MRI and
  wavelet transform
\newblock \emph{Neuroimage}
\newblock 62 (2012): 48--58.

\bibitem[\protect\citeauthoryear{Hazra, Reich, Reich, Shinohara, and
  Staicu}{Hazra et~al.}{2017}]{Hazra}
A.~Hazra, B.~J. Reich, D.~Reich, R.~Shinohara, and A.~Staicu.
\newblock A Spatio-Temporal Model for Longitudinal Image-on-Image Regression
\newblock \emph{Statistics in Biosciences}
\newblock  (2017): 1--25.

\bibitem[\protect\citeauthoryear{Higdon, Swall, and Kern}{Higdon
  et~al.}{1999}]{higdon}
D.~Higdon, J.~Swall, and J.~Kern.
\newblock Non-Stationary Spatial Modeling
\newblock \emph{Bayesian Statistics 6 - Proceedings of the Sixth Valencia
  Meeting}
\newblock  (1999): 761–768.

\bibitem[\protect\citeauthoryear{Jhuang, Fuentes, Jones, Esteves, Fancher,
  Furman, and Reich}{Jhuang et~al.}{2018}]{jhuang}
A.~Jhuang, M.~Fuentes, J.~Jones, G.~Esteves, C.~Fancher, M.~Furman, and
  B.~Reich.
\newblock Spatial Signal Detection Using Continuous Shrinkage Priors
\newblock \emph{In revision, Technometrics}
\newblock  (2018).

\bibitem[\protect\citeauthoryear{Jog, Carass, and Prince}{Jog
  et~al.}{2013}]{image4}
A.~Jog, A.~Carass, and J.~Prince.
\newblock Magnetic resonance image example-based contrast synthesis.
\newblock \emph{IEEE Trans Med Imaging}
\newblock 32 (2013): 2348–2363.

\bibitem[\protect\citeauthoryear{Jog, Carass, Roy, Pham, and Prince}{Jog
  et~al.}{2015}]{image2}
A.~Jog, A.~Carass, S.~Roy, D.~Pham, and J.~Prince.
\newblock Mr image synthesis by contrast learning on neighborhood ensembles
\newblock \emph{IEEE Trans Med Imaging}
\newblock 24 (2015): 63–76.

\bibitem[\protect\citeauthoryear{Jog, Carass, Roy, Pham, and Prince}{Jog
  et~al.}{2017}]{image3}
A.~Jog, A.~Carass, S.~Roy, D.~Pham, and J.~Prince.
\newblock Random forest regression for magnetic resonance image synthesis
\newblock \emph{IEEE Trans Med Imaging}
\newblock 35 (2017): 475–488.

\bibitem[\protect\citeauthoryear{Jones and Rice}{Jones and
  Rice}{1992}]{jones1992displaying}
M.~Jones and J.~A. Rice.
\newblock Displaying the important features of large collections of similar
  curves
\newblock \emph{The American Statistician}
\newblock 46 (1992): 140--145.

\bibitem[\protect\citeauthoryear{Kang, Reich, and Staicu}{Kang
  et~al.}{2016}]{kang}
J.~Kang, B.~J. Reich, and A.~M. Staicu.
\newblock Scalar-on-Image Regression via the Soft-Thresholded Gaussian Process
\newblock \emph{arXiv: 1604:03192}
\newblock  (2016).

\bibitem[\protect\citeauthoryear{Li, Zhang, Wang, Gonzalez, Maresh, and
  Coan}{Li et~al.}{2015}]{Li}
F.~Li, T.~Zhang, Q.~Wang, M.~Gonzalez, E.~Maresh, and J.~Coan.
\newblock Spatial Bayesian variable selection and grouping in high-dimensional
  scalar-on-image regressions
\newblock \emph{Annals of Applied Statistics}
\newblock 9 (2015): 687–713.

\bibitem[\protect\citeauthoryear{Mardia}{Mardia}{1970}]{Mardia}
K.~Mardia.
\newblock Measures of multivariate skewness and kurtosis with applications
\newblock \emph{Biometrika}
\newblock 57 (1970): 519--530.

\bibitem[\protect\citeauthoryear{Mejia, Sweeney, Dewey, Nair, Sati, Shea,
  Reich, and Shinohara}{Mejia et~al.}{2016}]{mejia}
A.~Mejia, E.~Sweeney, B.~Dewey, G.~Nair, P.~Sati, C.~Shea, D.~Reich, and
  R.~Shinohara.
\newblock Statistical estimation of $T_1$ relaxation times using conventional
  magnetic resonance imaging
\newblock \emph{Neuroimage}
\newblock 133 (2016): 176--188.

\bibitem[\protect\citeauthoryear{Mitchell and Beauchamp}{Mitchell and
  Beauchamp}{1988}]{spikeslab}
T.~J. Mitchell and J.~J. Beauchamp.
\newblock Bayesian variable selection in linear regression
\newblock \emph{Journal of the American Statistical Association}
\newblock 83 (1988): 1023–1036.

\bibitem[\protect\citeauthoryear{Morris, Baladandayuthapani, Herrick, Sanna,
  and Gutstein}{Morris et~al.}{2011}]{image1}
J.~Morris, V.~Baladandayuthapani, R.~Herrick, P.~Sanna, and H.~Gutstein.
\newblock Automated analysis of quantitative image data using isomorphic
  functional mixed models, with application to proteomics data
\newblock \emph{The Annals of Applied Statistics}
\newblock 5 (2011): 894–923.

\bibitem[\protect\citeauthoryear{Musgrove, Hughes, and Eberly}{Musgrove
  et~al.}{2017}]{fmri2}
D.~R. Musgrove, H.~Hughes, and L.~E. Eberly.
\newblock Fast, fully Bayesian spatiotemporal inference for fMRI data
\newblock \emph{Biostatistics}
\newblock  (2017).

\bibitem[\protect\citeauthoryear{Noh and Park}{Noh and Park}{2010}]{vcii1}
H.~Noh and B.~Park.
\newblock Sparse varying coefficient models for longitudinal data
\newblock \emph{Statistica Sinica}
\newblock 20 (2010): 1183--1202.

\bibitem[\protect\citeauthoryear{Nychka, Bandyopadhyay, Hammerling, Lindgren,
  and Sain}{Nychka et~al.}{2015}]{nychka}
D.~Nychka, S.~Bandyopadhyay, D.~Hammerling, F.~Lindgren, and S.~Sain.
\newblock A multiresolution Gaussian process model for the analysis of large
  spatial datasets
\newblock \emph{Journal of Computational and Graphical Statistics}
\newblock 24 (2015): 579--599.

\bibitem[\protect\citeauthoryear{Pomann, Staicu, Lobaton, Mejia, Dewey, Reich,
  Sweeney, and Shinohara}{Pomann et~al.}{2016}]{pomann}
G.~Pomann, A.~Staicu, E.~Lobaton, A.~Mejia, B.~Dewey, D.~Reich, E.~Sweeney, and
  R.~Shinohara.
\newblock A lag functional linear model for prediction of magnetization
  transfer ratio in multiple sclerosis lesions
\newblock \emph{The Annals of Applied Statistics}
\newblock 10 (2016): 2325--2348.

\bibitem[\protect\citeauthoryear{Reich, Guinness, Vandekar, Shinohara, and
  Staicu}{Reich et~al.}{2018}]{specimage}
B.~Reich, J.~Guinness, S.~Vandekar, R.~Shinohara, and A.~Staicu.
\newblock Fully-Bayesian spectral methods for imaging data
\newblock \emph{Biometrics}
\newblock  (2018).

\bibitem[\protect\citeauthoryear{Reich, Chang, and Foley}{Reich
  et~al.}{2014}]{reich2014spectral}
B.~J. Reich, H.~H. Chang, and K.~M. Foley.
\newblock A spectral method for spatial downscaling
\newblock \emph{Biometrics}
\newblock 70 (2014): 932--942.

\bibitem[\protect\citeauthoryear{Roberts and Rosenthal}{Roberts and
  Rosenthal}{1998}]{MHG}
G.~O. Roberts and J.~S. Rosenthal.
\newblock Optimal scaling of discrete approximations to Langevin diffusions
\newblock \emph{Journal of the Royal Statistical Society: Series B (Statistical
  Methodology)}
\newblock 60 (1998): 255--268.

\bibitem[\protect\citeauthoryear{Shen and Ghosal}{Shen and
  Ghosal}{2015}]{shen2015adaptive}
W.~Shen and S.~Ghosal.
\newblock Adaptive Bayesian procedures using random series priors
\newblock \emph{Scandinavian Journal of Statistics}
\newblock 42 (2015): 1194--1213.

\bibitem[\protect\citeauthoryear{Shinohara, Crainiceanu, Caffo, Gait\'an, and
  Reich}{Shinohara et~al.}{2011}]{shi2}
R.~Shinohara, C.~Crainiceanu, B.~Caffo, M.~Gait\'an, and D.~Reich.
\newblock Population-wide principal component-based quantification of
  blood-brain-barrier dynamics in multiple sclerosis
\newblock \emph{NeuroImage}
\newblock 57 (2011): 1430--1446.

\bibitem[\protect\citeauthoryear{Shinohara, Sweeney, Goldsmith, Shiee, and et.
  al.}{Shinohara et~al.}{2014}]{shi1}
R.~Shinohara, E.~Sweeney, J.~Goldsmith, N.~Shiee, and et. al.
\newblock Statistical normalization techniques for magnetic resonance imaging
\newblock \emph{NeuroImage}
\newblock 6 (2014): 9--19.

\bibitem[\protect\citeauthoryear{Stojanac, Suess, and Kliesch}{Stojanac
  et~al.}{2017}]{prodN}
Z.~Stojanac, D.~Suess, and M.~Kliesch.
\newblock On the distribution of a product of N Gaussian random variables
\newblock \emph{Wavelets and Sparsity XVII}
\newblock 10394 (2017): 1039419.

\bibitem[\protect\citeauthoryear{Stroud, Stein, and Lysen}{Stroud
  et~al.}{2016}]{Stroud}
J.~R. Stroud, M.~Stein, and S.~Lysen.
\newblock Bayesian and maximum likelihood Estimation for Gaussian processes on
  an incomplete lattice
\newblock \emph{Journal of computational and Graphical Statistics}
\newblock  (2016).

\bibitem[\protect\citeauthoryear{Sweeney, Shinohara, Shea, Reich, and
  Crainiceanu}{Sweeney et~al.}{2013}]{image5}
E.~Sweeney, R.~Shinohara, C.~Shea, D.~Reich, and C.~Crainiceanu.
\newblock Automatic lesion incidence estimation and detection in multiple
  sclerosis using multisequence longitudinal mri.
\newblock \emph{American Journal of Neuroradiology}
\newblock 34 (2013): 68–73.

\bibitem[\protect\citeauthoryear{Sweeney, Shinohara, Dewey, Schindler,
  Muschelli, Reich, Crainiceanu, and Eloyan}{Sweeney et~al.}{2016}]{sweeney}
E.~M. Sweeney, R.~T. Shinohara, B.~E. Dewey, M.~K. Schindler, J.~Muschelli,
  D.~S. Reich, C.~M. Crainiceanu, and A.~Eloyan.
\newblock Relating multi-sequence longitudinal intensity profiles and clinical
  covariates in incident multiple sclerosis lesions
\newblock \emph{NeuroImage}
\newblock 10 (2016): 1--17.

\bibitem[\protect\citeauthoryear{Tang, Wang, and Zhu}{Tang
  et~al.}{2013}]{vcii2}
Y.~Tang, H.~Wang, and Z.~Zhu.
\newblock Variable selection in quantile varying coefficient models with
  longitudinal data
\newblock \emph{Computational Statistics \& Data Analysis}
\newblock 57 (2013): 435--449.

\bibitem[\protect\citeauthoryear{Tibshirani}{Tibshirani}{1996}]{lasso}
R.~Tibshirani.
\newblock Regression Shrinkage and Selection via the Lasso
\newblock \emph{Journal of the Royal Statistical Society B}
\newblock 58 (1996): 267--288.

\bibitem[\protect\citeauthoryear{Tibshirani, Saunders, Rosset, Zhu, and
  Knight}{Tibshirani et~al.}{2005}]{fused}
R.~Tibshirani, M.~Saunders, S.~Rosset, J.~Zhu, and K.~Knight.
\newblock Sparsity and smoothness via the fused lasso
\newblock \emph{Journal of the Royal Statistical Society: Series B (Statistical
  Methodology)}
\newblock 67 (2005): 91--108.

\bibitem[\protect\citeauthoryear{Torabi, Moradzadeh, Vaziri, Ardekani, and
  Fatemizadeh}{Torabi et~al.}{2007}]{torabi2007multiple}
M.~Torabi, H.~Moradzadeh, R.~Vaziri, R.~D. Ardekani, and E.~Fatemizadeh.
\newblock Multiple sclerosis diagnosis based on analysis of subbands of 2-D
  wavelet transform applied on MR-images
\newblock  \emph{2007 IEEE/ACS International Conference on Computer Systems and
  Applications}.
\newblock 2007,   717--721.

\bibitem[\protect\citeauthoryear{Wang, Zhan, Chen, Zhang, Yang, Lu, Wang, Liu,
  and Phillips}{Wang et~al.}{2016}]{wang2016multiple}
S.-H. Wang, T.-M. Zhan, Y.~Chen, Y.~Zhang, M.~Yang, H.-M. Lu, H.-N. Wang,
  B.~Liu, and P.~Phillips.
\newblock Multiple sclerosis detection based on biorthogonal wavelet transform,
  RBF kernel principal component analysis, and logistic regression
\newblock \emph{IEEE Access}
\newblock 4 (2016): 7567--7576.

\bibitem[\protect\citeauthoryear{Wang and Zhu}{Wang and Zhu}{2017}]{Wang}
X.~Wang and H.~Zhu.
\newblock Generalized Scalar-on-Image Regression Models via Total Variation
\newblock \emph{Journal of the American Statistical Association}
\newblock 112 (2017): 1156--1168.

\bibitem[\protect\citeauthoryear{Wood and Chan}{Wood and Chan}{1994}]{wood}
A.~Wood and G.~Chan.
\newblock Simulation of stationary Gaussian process in $[0, 1]^d$
\newblock \emph{Journal of Computational and Graphical Statistics}
\newblock 3 (1994): 409--432.

\bibitem[\protect\citeauthoryear{Yan and Liu}{Yan and Liu}{2017}]{SIC}
B.~Yan and Y.~Liu.
\newblock Smooth Image-on-Scalar Regression for Brain Mapping
\newblock \emph{arXiv:1703.05264}
\newblock  (2017).

\bibitem[\protect\citeauthoryear{Zhang, Guindani, Versace, and et~al.}{Zhang
  et~al.}{2016}]{fmri1}
L.~Zhang, M.~Guindani, F.~Versace, and et~al.
\newblock A spatiotemporal nonparametric Bayesian model of multi-subject fMRI
  data
\newblock \emph{The Annals of Applied Statistics}
\newblock 10 (2016): 638–666.

\bibitem[\protect\citeauthoryear{Zhang, Lu, Zhou, Yang, Wu, Liu, Phillips, and
  Wang}{Zhang et~al.}{2016}]{zhang2016comparison}
Y.~Zhang, S.~Lu, X.~Zhou, M.~Yang, L.~Wu, B.~Liu, P.~Phillips, and S.~Wang.
\newblock Comparison of machine learning methods for stationary wavelet
  entropy-based multiple sclerosis detection: decision tree, k-nearest
  neighbors, and support vector machine
\newblock \emph{Simulation}
\newblock 92 (2016): 861--871.

\end{thebibliography}
\label{lastpage}
\end{document}